\documentclass[10pt,final]{elsart}
\usepackage{amsmath,amssymb}
\usepackage[pdftex]{graphicx}

\newcommand{\rmd}{\mathrm{d}}
\newcommand{\rme}{\mathrm{e}}
\newcommand{\rmi}{\mathrm{i}}
\newcommand{\ti}{t_{\mathrm{i}}}
\newcommand{\tf}{t_{\mathrm{f}}}
\newcommand{\Nt}{N_{\mathrm{t}}}

\newcommand{\Dt}{\Delta t}
\newcommand{\Dth}{\Delta\theta}
\newcommand{\bk}{\boldsymbol{k}}

\newcommand{\calO}{\mathcal{O}}
\newcommand{\calD}{\mathcal{D}}
\newcommand{\unity}{\mathrm{I}}

\newcommand{\sfD}{\mathsf{D}}
\newcommand{\phir}{\phi_{\mathrm R}}
\newcommand{\phii}{\phi_{\mathrm I}}

\def\hsymb#1{\mbox{\strut\rlap{\smash{\Huge$#1$}}\quad}}
\def\lsymb#1{\mbox{\strut\rlap{\smash{\Large$#1$}}\quad}}

\begin{document}
\begin{frontmatter}
\begin{flushright}
RIKEN-QHP-150
\end{flushright}

\title{Restricted phase-space approximation\\
 in real-time stochastic quantization}

\author[tokyo2]{Ryoji Anzaki}
\author[tokyo]{Kenji Fukushima}
\author[riken]{Yoshimasa Hidaka}\
and
\author[tokyo2]{Takashi Oka}

\address[tokyo2]{Department of Applied Physics, The University of Tokyo,
                 7-3-1 Hongo, Bunkyo-ku, Tokyo 113-0033, Japan}
\address[tokyo]{Department of Physics, The University of Tokyo,
                7-3-1 Hongo, Bunkyo-ku, Tokyo 113-0033, Japan}
\address[riken]{Theoretical Research Division, Nishina Center,
                RIKEN, Wako 351-0198, Japan}

\begin{abstract}
 We perform and extend real-time numerical simulation of a
 low-dimensional scalar field theory or a quantum mechanical system
 using stochastic quantization.  After a brief review of the
 quantization method and the complex Langevin dynamics, we calculate
 the propagator and make a comparison with analytical results.  This
 is a first step toward general applications, and we focus only on the
 vacuum properties of the theory;  this enables us to handle the
 boundary condition with the $\rmi\epsilon$ prescription in frequency
 space.  While we can control stability of the numerical simulation
 for any coupling strength, our results turn out to flow into an
 unphysical fixed-point, which is qualitatively understood from the
 corresponding Fokker-Planck equation.  We propose a simple truncation
 scheme, ``restricted phase-space approximation,'' to avoid the
 unphysical fixed-point.  With this method, we obtain stable results
 at reasonably good accuracy.  Finally we give a short discussion on
 the closed-time path formalism and demonstrate the direct computation
 of the vacuum expectation value not with the $\rmi\epsilon$
 prescription but from an explicit construction of the Feynman
 kernel.
\end{abstract}
\end{frontmatter}

\section{Introduction}

Large-scale numerical computation is becoming a vital building block
in today's scientific researches.  In theoretical physics, the
numerical approach is regarded as a starting point of a pursuit toward
fundamental understanding of new phenomena.  Performing numerical
\textit{experiments}, we can test ideas and hypotheses in an ideal
setup repeatedly and easily, which is usually difficult in real
experiments.  This enables us to efficiently build models and theories
that describe nature.  In this spirit, in order to study new physics,
it is important to develop new numerical methods and extend their
validity.

Quantum field theories that accommodate infinite degrees of freedom
stand in the center of modern physics.  It is becoming less and less
costly to perform large-scale numerical simulations thanks to
tremendous developments in the computing power and the various
innovations in the numerical algorithms.  One area where computers are
playing an important role is the fundamental theory of the strong
interaction; that is, quantum chromodynamics
(QCD)~\cite{Brambilla:2014aaa} can be formulated on the
four-dimensional lattice grid in Euclidean space-time, so that the
exponentiated action, $\rme^{-S_{\rm QCD}}$, is a real positive number
and can be interpreted as a weight factor in analogy with statistical
mechanics~\cite{Gupta:1997nd}.  We can then carry out the functional
integral by means of the Monte-Carlo algorithm as long as the weight
factor is real and non-negative.  This approach known as the
lattice-QCD simulation~\cite{rothe2012lattice} has been the most
successful non-perturbative tool to investigate the QCD-vacuum
(topological) structure~\cite{D'Elia:2012vv}, thermodynamics of QCD
matter~\cite{Bazavov:2011nk,Borsanyi:2013bia}, the hadron
spectroscopy~\cite{Fodor:2012gf}, and also the real-time characters
such as the spectral function~\cite{Nakahara:1999vy,Asakawa:2000tr},
the particle production rate~\cite{Karsch:2001uw,Ding:2010ga}, and the
transport coefficients~\cite{Nakamura:2004sy,Meyer:2007ic,%
Meyer:2007dy,Aarts:2007wj,Amato:2013naa}, etc.

Another area where numerical simulations are intensively utilized is
condensed matter physics, especially in the field of strongly
correlated electron systems.  It has been realized that quantum
many-body effect leads to various phase transitions.  A well studied
example is the Mott transition~\cite{Imada98} in which electrons
freeze their motion due to strong Coulomb interaction.  It is believed
that this transition is relevant to understanding of the pairing
mechanism of high temperature superconductivity~\cite{Nagaosa06}.
Numerical algorithms such as the density matrix renormalization group
(DMRG)~\cite{White92,Schollwock05} and the dynamical mean field theory
(DMFT)~\cite{Georges96} have been developed and applied to problems in
correlated electron systems.  Recently, real-time dynamics in
condensed matter systems is becoming a hot topic (for a review see
Ref.~\cite{Aoki14}).  One important problem is the physics of quantum
quench; i.e., many-body dynamics triggered by a rapid parameter
change.  Phenomena such as ``prethermalization''~\cite{Berges:2004ce},
initially discovered in the QCD community, has motivated 
many condensed matter researches~\cite{Moeckel2010a,Eckstein2010}.  In
condensed matter, variety of methods exist to deal with the problem of
a time-evolving quantum many-body system.  They range from direct
wave-function-based techniques such as exact diagonalization and
DMRG~\cite{tddmrg,Daley2004,White2004a}, quantum master
equations~\cite{BreuerPetruccioneBook}, and quantum kinetic
equations~\cite{RammerBook}, to the Keldysh formalism for the
non-equilibrium Green¡Çs
functions~\cite{kadanoff1962quantum,Keldysh:1964ud,Schwinger:1960qe}. 
Using the Keldysh formalism, many sophisticated theoretical techniques
that were developed for equilibrium can be straightforwardly utilized
to non-equilibrium systems; e.g., diagrammatic quantum
Monte-Carlo method (QMC)~\cite{Werner09,Muehlbacher2008} and
non-equilibrium DMFT~\cite{Aoki14,Freericks2006}.  The price to pay is
the severe negative sign problem, and it is still challenging to study
the long time behavior.  We summarize major approaches for
non-equilibrium many-body systems in Table~\ref{table1}.

\begin{table}[tb]
\begin{center}
    \begin{tabular}{ | l || l | l | l | }
    \hline
    Method & Quantum & Variables & Limitation   \\ \hline\hline
    Stochastic quantization & Full & Fields $\phi(x,t,\theta)$
      & Unphysical fixed-point  \\ \hline
    Classical statistical sim. & $\calO(\hbar)$
      & Fields $\phi(x,t)$ & Large occupation num.  \\ \hline
    Real-time QMC & Full & Green's func. & Sign problem  \\ \hline
    Time-dependent DMRG & Full & Wave function & Low-dim
      systems \\ \hline
    Non-equilibrium DMFT & Full & Green's func. &
      Short time  \\ \hline
    \end{tabular}
\end{center}
\vspace{0.7em}
\caption{Numerical methods for real-time calculations.  The classical
  statistical simulation contains quantum fluctuations only up to
  $\calO(\hbar)$ but the long-time simulation is possible, while other
  methods are fully quantum.  Each method has an advantage and a
  limitation of the validity as listed.}
\label{table1}
\vspace{1em}
\end{table}

In order to study non-linear QCD processes far from equilibrium such
as the pattern formation~\cite{Cross:1993} and the turbulent
flow~\cite{univis91164995}, a method that can treat not only fermions
but also bosons must be developed.  To overcome the limitation of the
Monte-Carlo simulation, some alternative approaches are proposed such
as the gauge/gravity
correspondence~\cite{Aharony:1999ti}, the classical statistical field
theory~\cite{Polarski:1995jg,Son:1996zs,Khlebnikov:1996mc}, the
2-particle-irreducible formalism~\cite{Cornwall:1974vz} (see also
Ref.~\cite{kadanoff1962quantum}), and the stochastic
quantization~\cite{Nelson:1966sp,Parisi:1980ys,%
Damgaard:1987rr,Namiki:1993fd}.

The gauge/gravity correspondence has provided us with useful insights
into the thermalization problem and the numerical simulations are
possible now to trace the evolution processes of the dynamical
system~\cite{Chesler:2008hg,Chesler:2009cy,Grumiller:2008va,%
Balasubramanian:2010ce,Balasubramanian:2011ur,Heller:2011ju,%
Hashimoto:2010wv}, though the technique can be applied only to a
special class of the strong-coupling gauge theory.  The classical
statistical simulation, which is also known as the ``truncated
Wigner'' approximation~\cite{Polkovnikov:2010}, is quite successful in
describing the early stages of the relativistic heavy-ion
collision~\cite{Romatschke:2005pm,Romatschke:2006nk,Fukushima:2006ax,%
Gelis:2007kn}, which has been closely investigated in connection to
the wave turbulence and the scaling behavior
also~\cite{Dusling:2010rm,Epelbaum:2011pc,Dusling:2012ig,%
Fukushima:2011nq,Fukushima:2013dma,Berges:2013eia,Berges:2013fga,%
Berges:2013lsa,Kasper:2014uaa}.

Although the classical statistical simulation is a useful tool in the
regime where the occupation number is large enough to justify the
classical treatment, the formalism itself needs to be elaborated not
to ruin the renormalizability~\cite{Epelbaum:2014yja}.  For this
purpose it is an interesting question to think of a possible relation
between the classical statistical approach and stochastic quantization
as speculated in Ref.~\cite{Fukushima:2014iqa}, that has been hinted
also by the simulation in Ref.~\cite{Gelis:2013oca}.  Needless to say,
if one can perform a direct real-time simulation with stochastic
quantization without making any approximation, we can go beyond the
limitation $\calO(\hbar)$ of the classical statistical approximation.
It should be intriguing to pursue such an ultimate goal.

There have been several attempts to solve the real-time theories using
stochastic quantization
numerically~\cite{Berges:2005yt,Berges:2006xc,Berges:2007nr}, which,
however, did not succeed in proceeding far out of equilibrium.  As we
will explain later, we should then solve a diffusion equation with a
pure-imaginary coefficient together with stochastic random variables;
i.e., a complex Langevin equation~\cite{Parisi:1984cs,Huffel:1984mq}.
We are often stuck with two major obstacles in handling the complex
Langevin equation:  one is the numerical instability, and the other is
the problem of run-away trajectories (i.e., physical instability).
Not only in the context of real-time physics, but also in the efforts
to attack the so-called sign problem at finite
density~\cite{Muroya:2003qs}, the adaptive step-size method is
developed to suppress the numerical instability and the convergence
is under careful investigation~\cite{Aarts:2013bla,Aarts:2013uxa}.
The stochastic quantization method has also been utilized in the
application of the Lefschetz thimble to evade the sign
problem~\cite{Cristoforetti:2012su,Fujii:2013sra,%
Cristoforetti:2013qaa,Mukherjee:2014hsa,Cristoforetti:2014gsa}.

Because the theoretical interest in the potential of stochastic
quantization is growing lately in various research fields, it is quite
timely to revisit this method to perform a direct real-time
simulation.  In this paper we do not assume that the initial state is
in thermal equilibrium (which will enhance stability of the
simulation~\cite{Berges:2006xc}) but limit ourselves to the vacuum
properties only, for which the information on the initial and final
wave-functionals are to be dropped by the $\rmi\epsilon$
prescription.  Besides, we can check if our numerical results are on
the right physical trajectory or not as long as the vacuum properties
are somehow known.  Our ultimate goal shall be the study of full
quantum and non-equilibrium phenomena, and in the final section, we
will briefly sketch an outlook along these lines.

\section{Scalar Field Theory in Minkowski Space-time}
\label{sec:real}

To make our discussions self-contained, we shall make a brief overview
of stochastic quantization here for a real scalar field theory (see
reviews~\cite{Damgaard:1987rr,Namiki:1993fd} for more details).
Readers who are familiar with real-time formalism and stochastic
quantization can skip this section and jump to
Sec.~\ref{sec:numerical}.

\subsection{Formalism}

\begin{figure}
 \begin{center}
 \includegraphics[width=0.75\textwidth]{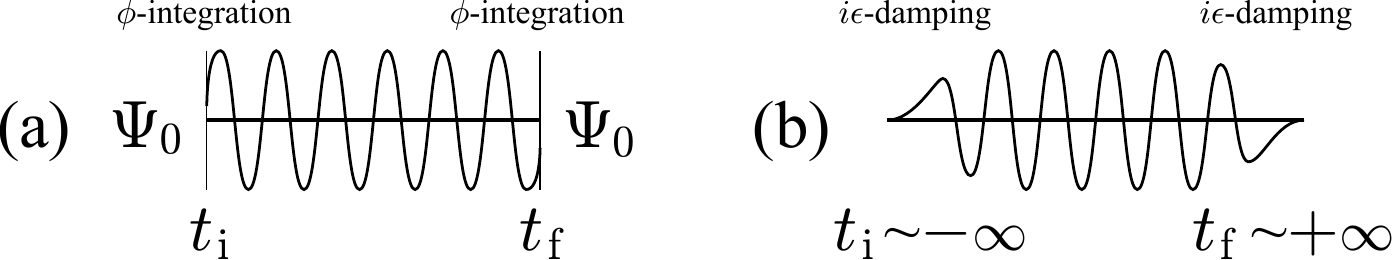}
 \end{center}
 \caption{Schematic illustration of the vacuum amplitude in two
   different but equivalent ways:  (a) The convolution with the vacuum
   wave-functional $\Psi_0$ is taken at $\ti$ and $\tf$.  (b) The
   boundary is irrelevant for sufficiently large $\ti$ and $\tf$ as a
   result of the damping by $\rmi\epsilon$.}
 \label{fig:schematic}
\vspace{5mm}
\end{figure}

Let us begin with the general formulation of quantum field theory.  We
denote the amplitude from the initial $|\Psi_{\rm i},\ti\rangle$ to
the finial $|\Psi_{\rm f},\tf\rangle$ as
$\langle \Psi_{\rm f},\tf|\Psi_{\rm i},\ti\rangle$, which we can
rewrite in the functional integral form as follows:
\begin{equation}
 \langle \Psi_{\rm f},\tf|\Psi_{\rm i},\ti\rangle
 = \langle \Psi_{\rm f}|\,\rme^{-\rmi H(\tf-\ti)}| \Psi_{\rm i}\rangle
 = \int\!\calD \phi\, \rme^{\rmi S[\phi]}\,
  \Psi_{\rm f}^\ast[\phi(\tf)]\,\Psi_{\rm i}[\phi(\ti)]
\label{eq:pathIntegral}
\end{equation}
with $H$ being the Hamiltonian.  Then, the $n$-point Green's functions
read:
\begin{equation}
 \begin{split}
  & G^{(n)}(x_1,x_2,\cdots,x_n;\tf;\Psi_{\rm f},\tf;\Psi_{\rm i},\ti) \\
  &\qquad \equiv\frac{1}{\langle \Psi_{\rm f},\tf|\Psi_{\rm i},\ti\rangle}
   \langle \Psi_{\rm f},\tf| T\,\phi(x_1)\phi(x_2)\cdots \phi(x_n)
   |\Psi_\mathrm{i},\ti\rangle\\
  &\qquad=\frac{\displaystyle \int \!\calD \phi\,
   \rme^{\rmi S[\phi]} \,\Psi_{\rm f}^\ast[\phi(\tf)] \,
   \Psi_{\rm i}[\phi(\ti)] \,\phi(x_1)\,\phi(x_2)\cdots \phi(x_n)}
   {\displaystyle \int\!\calD \phi\, \rme^{\rmi S[\phi]} \,
   \Psi_{\rm f}^\ast[\phi(\tf)] \, \Psi_{\rm i}[\phi(\ti)]} \;,
\label{eq:GreenFunction}
\end{split}
\end{equation}
where $T$ denotes the time-ordered-product operator.  These general
Green's functions obviously depend on the choice of the initial and
finial wave functionals.  Thus, for the calculation of the amplitude,
the real-time evolution of quantum systems is formulated as the
boundary problem rather than the initial-value problem as in classical
physics.  We are sometimes interested in the vacuum properties also,
which can be accessed either by convoluting the vacuum wave-functional
$\Psi_0[\phi]$ (see Eq.~\eqref{eq:gaussian} for an explicit form and
Fig.~\ref{fig:schematic}~(a) for an illustration) or by taking
$\tf-\ti\to\infty$ with Feynman's $\rmi\epsilon$ prescription (see
Fig.~\ref{fig:schematic}~(b)): $H\to H(1-\rmi\epsilon)$.  Inserting
the complete set onto
$\rme^{-\rmi H(\tf-\ti)(1-\rmi\epsilon)}|\Psi_{\rm i}\rangle$, we can
extract the dominant contribution in this limit as
\begin{equation}
\begin{split}
 \rme^{-\rmi H(\tf-\ti)(1-\rmi\epsilon)}|\Psi_{\rm i}\rangle
  &= \sum_n \rme^{-\rmi E_n(\tf-\ti)(1-\rmi\epsilon)}|n\rangle\langle n
   |\Psi_{\rm i}\rangle \\
  &= \rme^{-\rmi E_0(\tf-\ti)(1-\rmi\epsilon)}|\Omega\rangle
   \langle\Omega|\Psi_{\rm i}\rangle \; \Bigl[ 1 +
   \calO(\rme^{-(E_1-E_0)(\tf-\ti)\epsilon}) \Bigr] \;.
\end{split}
\end{equation}
Thus, the vacuum state $|\Omega\rangle$ dominates in the presence of
small but finite $\epsilon$.  In this case, the Green's functions
given in Eq.~\eqref{eq:GreenFunction} become insensitive to any
excited states but the vacuum state; i.e.,
\begin{equation}
 G^{(n)}(x_1,x_2,\cdots) = \langle\Omega|T\phi(x_1)\phi(x_2)
  \cdots \phi(x_n)|\Omega\rangle\;,
\end{equation}
where the normalization of the vacuum is assumed to be
$\langle\Omega|\Omega\rangle=1$.  In the numerical simulation,
practically, $\tf-\ti$ cannot be infinity, and thus we need to keep
$(E_1-E_0)(\tf-\ti)\epsilon\gg 1$ to make the vacuum state dominate
over any excited states.  We present a schematic illustration in
Fig.~\ref{fig:schematic} to sketch these two alternative methods to
extract the vacuum amplitude.

\begin{figure}
 \begin{center}
 \includegraphics[width=\textwidth]{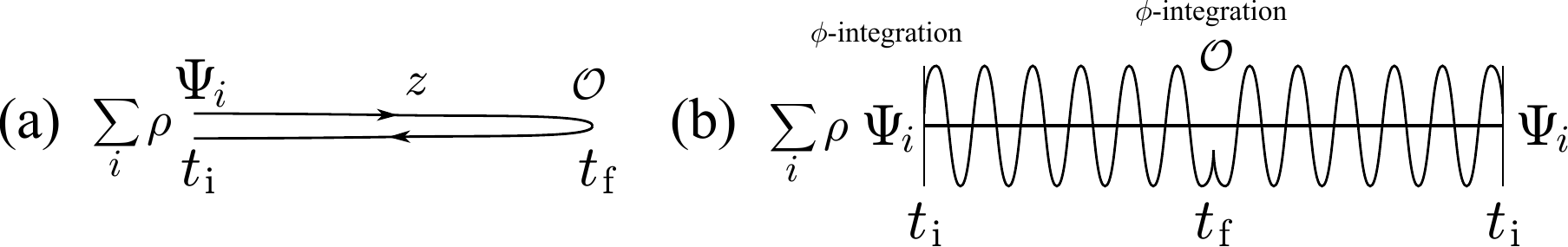}
 \end{center}
 \caption{Schematic illustration of the expectation value.  (a)
   Representation in the closed-time path formalism by introducing a
   complexified time variable $z$.  (b) Unfolded representation
   analogous to the amplitude calculations as depicted in
   Fig.~\ref{fig:schematic}.} 
 \label{fig:schematic2}
\vspace{5mm}
\end{figure}

For more general problems out of equilibrium we often need an
expectation value of some operator $\calO$ at $\tf$ with an initial
condition given at $\ti$, which we can express as
\begin{equation}
 \langle \calO \rangle_{\tf}
  \equiv \sum_{\Psi_{\rm i}} \langle\Psi_{\rm i};\ti|\,
  \rho\, \rme^{-\rmi H(\ti-\tf)}\, \calO\,
   \rme^{-\rmi H(\tf-\ti)}\, |\Psi_{\rm i};\ti\rangle \;.
\label{eq:ex_O}
\end{equation}
Here the density matrix $\rho$ specifies the initial state at
$\ti$.  Using a complexified time variable $z$, we can regard
Eq.~\eqref{eq:ex_O} as an ``amplitude'' computed on the closed-time
path; see Fig.~\ref{fig:schematic2}~(a).  If the system is thermal and
$\rho$ takes a form of $\rho=\rme^{-H/T}/(\mathrm{tr}\,\rme^{-H/T})$,
it would be an elegant representation of the theory if we combine all
time-evolution operators, $\rme^{-\rmi H(\tf-\ti)}$,
$\rme^{-\rmi H(\ti-\tf)}$, and $\rme^{-H/T}$ together with $z$ running
along a single path on the complex plane.  This is nothing but the
real-time formalism of the finite-temperature field
theory~\cite{Landsman:1986uw}.  In this manner we can recover the
well-known $2\times 2$ matrix structure of the propagator from a
combination of the forward ($\ti\to\tf$) and the backward path
($\tf\to\ti$).  The off-diagonal components pick up $\rho$, and in
the case of thermal equilibrium, they contain the thermal distribution
function.

For a general $\rho$, we can no longer incorporate $\rho$ as a
deformation of the time path, and we should close the time path with
an explicit insertion of the density matrix at initial $\ti$ as
sketched in Fig.~\ref{fig:schematic2}~(a).  This is the basic
description of the closed-time path (CTP) or the Schwinger-Keldysh
formalism~\cite{Keldysh:1964ud,Schwinger:1960qe}.  We note that the
closed-time path is often extended to $\tf=\infty$ for convenience
especially when the perturbative calculation is formulated.  Putting a
source $J(z)$ along the path of Fig.~\ref{fig:schematic2}~(a), we can
construct an arbitrary operator by taking $\delta/\delta J(z)$.

As we will discuss in great details in Sec.~\ref{sec:closed}, in
stochastic quantization, we do not have to introduce the $2\times2$
matrix propagator.  We can actually unfold the closed-time path as in
Fig.~\ref{fig:schematic2}~(b), so that we can perform direct numerical
simulations to evaluate the expectation value in the same way as the
amplitude calculation.  One may think that the time evolution for
$\tf\to\ti$ is just a duplicate of that for $\ti\to\tf$, and this is
true as we see in Sec.~\ref{sec:closed}, but this seemingly redundant
reflection plays an important role for the treatment of a singular
edge at $\tf$.

Let us now proceed to the formulation of stochastic quantization.  In
Euclidean field theories the weight appears in the functional
integral~\eqref{eq:pathIntegral} as a Boltzmann factor $\rme^{-S_E}$
with the Euclidean action $S_E$.  The stochastic process or the
corresponding Fokker-Planck equation can generate such a weight
correctly;  in other words, quantum fluctuations are encoded in a
form of the Langevin dynamics, which was proposed by Parisi and
Wu~\cite{Parisi:1980ys} and is commonly called ``stochastic
quantization.''   In Minkowski space-time, however, the weight factor
$\rme^{\rmi S}$ takes a complex value and the convergence with the
Fokker-Planck equation is a subtle problem, while the Langevin
dynamics reproduces the ordinary perturbative series.

The real scalar field theory of our present interest is defined with
the following action:
\begin{equation}
 S = \int \rmd^d x\, \biggl[ \frac{1}{2} (\partial_\mu\phi)
  (\partial^\mu\phi) - \frac{m^2}{2} \phi^2
   - \frac{\lambda}{4} \phi^4 \biggr] \;,
\end{equation}
where $d$ is the number of space-time dimensions and we consider the
$\phi^4$-interaction only.  The corresponding (complex) Langevin
equation turns out to be
\begin{equation}
\begin{split}
 \partial_\theta\,\phi(x,\theta) &= \rmi\,
  \frac{\delta S}{\delta\phi(x)} \biggr|_{\phi(x)\to\phi(x,\theta)}
  \!\!\! +\eta(x,\theta)\\
 & = -\rmi (\square + m^2 - \rmi\epsilon)\, \phi(x,\theta)
     -\rmi\lambda\,\phi^3(x,\theta) + \eta(x,\theta) \;,
\label{eq:Langevin}
\end{split}
\end{equation}
where $\theta$ is the fictitious time not related to physical
coordinates and it runs from $0$ to $\infty$ in a conventional
choice.  We denote the stochastic noise term by $\eta(x,\theta)$.

We should fix a starting condition at $\theta=0$ and the simplest
prescription is $\phi(x,\theta=0)=0$.  It is possible to choose a
non-zero initial condition, but it will be vanishing at
$\theta\to\infty$ and so irrelevant to the final results as long as we
utilize the $\rmi\epsilon$ prescription.  To recover the ordinary
perturbative expansion of the $\phi^4$-theory, the stochastic noise
should satisfy
\begin{equation}
 \langle\eta(x,\theta)\,\eta(x',\theta')\rangle_\eta
  = 2\,\delta^{(d)}(x-x')\,\delta(\theta-\theta') \;.
\label{eq:noise}
\end{equation}
In other words, the above expression gives us a definition of the
average procedure over $\eta(x,\theta)$ as a Gaussian average.  If we
want to know the vacuum expectation value of some operator
$\calO$, we should calculate the $\eta$-average of
$\calO[\phi(x,\theta)]$ where the $\eta$-dependence comes in through
the $\theta$-evolution of $\phi(x,\theta)$ according to
Eq.~\eqref{eq:Langevin}.  This means that
\begin{equation}
 \langle\calO[\phi(x)]\rangle = \lim_{\theta\to\infty}
  \langle\calO[\phi(x,\theta)]\rangle_\eta \;.
\end{equation}
Here, precisely speaking, the vacuum expectation value in the
left-hand side represents the time-ordered quantity as usual in the
functional integration formalism.

Now that we finish a quick flash of stochastic quantization, let us
make sure that it certainly produces the ordinary perturbation theory,
which also turns out to be useful for later discussions about the
numerical simulation.

\subsection{Recovery of the free propagator}

It is the most convenient to move to the Fourier space to solve the
complex Langevin equation~\eqref{eq:Langevin} analytically.  We define
the scalar field and stochastic variables in momentum space as
\begin{equation}
 \phi_k(\theta) \equiv \int\rmd^d x\, \phi(x,\theta)\,
  \rme^{\rmi k\cdot x} \;,\qquad
 \eta_k(\theta) \equiv \int\rmd^d x\, \eta(x,\theta)\,
  \rme^{\rmi k\cdot x}
\end{equation}
with the four-vector notation: $k\equiv(\omega,\bk)$.  We can then
recast the differential equation with $\lambda=0$ (i.e., free theory)
into the following form:
\begin{equation}
 \partial_\theta\,\phi_k(\theta) = \rmi (\omega^2
  -\xi_{\bk}^2 + \rmi\epsilon)\,\phi_k(\theta) + \eta_k(\theta) \;,
\label{eq:diff_eq_w}
\end{equation}
where the stochastic noise in this Fourier transformed basis is
characterized by the following average:
\begin{equation}
 \langle\eta_k(\theta)\,\eta_{k'}(\theta')\rangle_\eta
  = 2\,(2\pi)^{d}\delta(\omega+\omega')\,\delta^{(d-1)}(\bk+\bk')\,
    \delta(\theta-\theta') \;.
\end{equation}

It is a simple exercise to find an analytical solution of this linear
differential equation of Eq.~\eqref{eq:diff_eq_w} that yields
\begin{equation}
 \begin{split}
 \phi_k(\theta) &= \frac{1}{\partial_\theta - \rmi
  (\omega^2 - \xi_{\bk}^2 + \rmi\epsilon)} \,\eta_k(\theta)
  + \rme^{\rmi(\omega^2 - \xi_{\bk}^2 +\rmi\epsilon)\theta} \phi_k(0)  \\
 & = \int_0^\theta \rmd\theta'\, \rme^{\rmi(\omega^2-\xi_{\bk}^2
  +\rmi\epsilon)(\theta-\theta')}\,\eta_k(\theta')
  + \rme^{\rmi(\omega^2 - \xi_{\bk}^2 +\rmi\epsilon)\theta} \phi_k(0) \;.
 \end{split}
\end{equation}
The propagator is a two-point function constructed with the above
$\phi_k$.  We should keep in mind to take the $\theta\to\infty$ limit
carefully after taking the two-point function.  The free Feynman
propagator is immediately obtainable through
\begin{align}
 G_0(k,k') &= \lim_{\theta\to\infty} \langle \phi_k(\theta)
  \phi_{k'}(\theta) \rangle_\eta \notag\\
 &= (2\pi)^{d}\delta(\omega+\omega')\,\delta^{(d-1)}(\bk+\bk')\,
  \frac{\rmi}{\omega^2-\xi_{\bk}^2+\rmi\epsilon}\,\lim_{\theta\to\infty}
  \Bigl[1-\rme^{2\rmi(\omega^2-\xi_{\bk}^2+\rmi\epsilon)\theta}\Bigr] \notag\\
  &\qquad\qquad\qquad
  + \lim_{\theta\to\infty}
  \rme^{\rmi(\omega^2+{\omega'}^2 - \xi_{\bk}^2- \xi_{\bk'}^2 +2\rmi\epsilon)\theta}
  \phi_k(0) \phi_{k'}(0) \;.
\label{eq:green}
\end{align}
It is important to note that we can safely take the $\theta\to\infty$
limit thanks to the presence of $\epsilon>0$.  In other words, this
$\rmi\epsilon$ term was needed in Eq.~\eqref{eq:diff_eq_w} for the
convergence in the $\theta\to\infty$ limit and such an insertion
is completely consistent with the well-known $\rmi\epsilon$
prescription to get the Feynman (time-ordered) propagator;  the second
oscillatory term inside of the square brackets and the last term in
Eq.~\eqref{eq:green} vanish, so that the standard expression of the
free Feynman propagator emerges.  The final result is independent of
the choice of initial wave-functional, and this is true for any
higher-order diagrams, so that we can freely adopt the initial
condition as $\phi_k(0)=0$ in the following.

For the direct real-time simulation, hence, we should keep a finite
$\epsilon$ in principle and integrate the complex Langevin
equation with respect to $\theta$ up to a sufficiently large value to
fulfill $\rme^{-2\epsilon\theta}\ll 1$.  However, it is impractical to
realize such a condition strictly.  We will come back to this point
when we present our numerical results later.

\subsection{Recovery of the perturbative expansion}
\label{sec:perturbative}

\begin{figure}
 \includegraphics[width=0.48\textwidth]{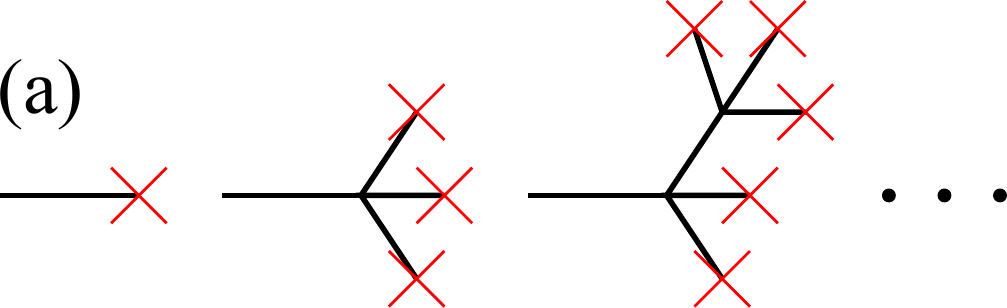} \hspace{2em}
 \includegraphics[width=0.41\textwidth]{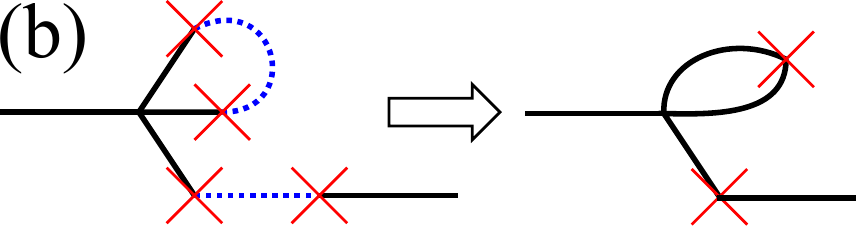}
 \caption{(a) Stochastic diagrams that represent an iterative solution
 of the integral equation~\eqref{eq:integral_eq}.  The crosses are the
 stochastic variables $\eta_k(\theta)$.\ \
 (b) An example of contraction of the stochastic variables for the
 two-point function that produces a Feynman diagram of the
 self-energy.}
 \label{fig:stochastic}
\vspace{5mm}
\end{figure}

With a finite $\lambda$ of the self-interaction strength, we cannot
write a full analytical solution down but still find a recursion
equation or an integral equation, from which we can iteratively
produce a solution of the differential equation.  That is, the complex
Langevin equation in momentum space translates into
\begin{equation}
 \begin{split}
 \phi_k(\theta) \!=\! \int_0^\theta \rmd\theta'\,
  \rme^{\rmi(\omega^2-\xi_{\bk}^2+\rmi\epsilon)(\theta-\theta')}\,
  \Bigl[ \eta_k(\theta') \!-\! \rmi\lambda \int
  \frac{\rmd^d k_1 \rmd^d k_2}{(2\pi)^{2d}}\,
  \phi_{k-k_1-k_2}(\theta')\phi_{k_1}(\theta')\phi_{k_2}(\theta')
  \Bigr] .
 \end{split}
\label{eq:integral_eq}
\end{equation}
This is a convenient expression used for the iteration that generates
the expansion of $\phi_k(\theta)$ in powers of $\lambda$.  The number
of involved $\eta_k(\theta)$ would increase as we go to higher-order
terms in the $\lambda$-expansion, which is graphically illustrated in
Fig.~\ref{fig:stochastic}~(a).

Because of the Gaussian nature of the stochastic
variables~\eqref{eq:noise}, the $\eta$-average makes a pair of
$\eta_k(\theta)$ contracted to each other.
Figure~\ref{fig:stochastic}~(b) shows an example of such contraction
in the computation of $\langle\phi_k(\theta)\phi_{k'}(\theta)\rangle$.
The dotted lines indicate the contracted pairs of $\eta_k(\theta)$ and
the contraction results in the lowest-order Feynman diagram of the
self-energy.  This procedure is readily generalized to higher-order
contributions, so that the perturbative series from the ordinary
quantization scheme are exactly
recovered~\cite{Huffel:1984mq,Grimus:1982wn,Huffel:1985ma}.

\section{Numerical Simulations without Interaction}
\label{sec:numerical}

First we shall focus on the simple case without interaction; i.e.,
$\lambda=0$ and for a fixed spatial momentum.  The system then reduces
to a 0+1 dimensional problem or a quantum mechanical problem.  We
reproduce the free propagator by numerical means of stochastic
quantization.  We explain the discretization schemes in momentum
(frequency) space, and then we address the numerical results.

\subsection{Discretization in frequency space}

Because the boundaries at $\ti$ and $\tf$ are irrelevant in the
$\rmi\epsilon$ prescription, we can implicitly impose the periodic
boundary condition so that we can work in frequency space (with the
spatial momentum $\bk$ frozen).  Then, we should solve
Eq.~\eqref{eq:diff_eq_w} numerically for a given $\xi$, where we drop
the subscript $\bk$.  So, the system in what follows has only
$t$-dependence with a mass scale given by $\xi$; in other words, this
is a harmonic oscillator problem in quantum mechanics.

On the lattice we discretize the frequency and the mass scale as
\begin{equation}
 \omega = \omega_{\rm min}\,\nu\;,\qquad
 \xi = \omega_{\rm min}\,\mu \qquad \text{with} \qquad
 \omega_{\rm min} = \frac{2\pi}{\Nt+1} \;,
\label{eq:munu}
\end{equation}
under the condition that the time $t$ runs from $\ti=0$ to
$\tf=\Nt\Dt$ with a period $(\Nt+1)\Dt$.  We sometimes drop the time
spacing $\Dt$ or express quantities in the unit of $\Dt$.  With
discretization we should generate the stochastic variables in such a
way as
\begin{equation}
 \langle\eta_\nu(\theta)\,\eta_{\nu'}(\theta')\rangle
  = \frac{2\Nt}{\Dth}\, \delta_{\nu+\nu',0}\,\delta_{\theta,\theta'}\;,
\end{equation}
where $\Dth$ is a lattice spacing in the $\theta$ direction.  We note
that the generation of $\eta_\nu(\theta)$ needs some caution to make
$\eta(x,\theta)$ in coordinate space non-complex, which demands:
$\eta_{-\nu}(\theta) = \eta_\nu^\ast(\theta)$.  Therefore, when we
generate $\eta_\nu(\theta)$, we first generate real stochastic variables
$\bar{\eta}_1$ and $\bar{\eta}_2$ and then combine them as
\begin{equation}
 \eta_\nu(\theta) = \bar{\eta}^{(1)} + \rmi \bar{\eta}^{(2)}\;,\qquad
 \eta_{-\nu}(\theta) = \bar{\eta}^{(1)} - \rmi \bar{\eta}^{(2)}
\end{equation}
for $\nu\neq 0$ and
\begin{equation}
 \eta_0(\theta) = \sqrt{2}\,\bar{\eta}^{(1)}
\end{equation}
for $\nu=0$.  The differential equations to be solved are thus
\begin{equation}
 \phi_\nu(\theta+\Dth) = \frac{\rme^{-\epsilon\Dth}}
  {1-\rmi\omega_{\rm min}^2(\nu^2-\mu^2)\Dth}\,\phi_\nu(\theta)
  + \Dth\, \eta_\nu(\theta)
\label{eq:diff}
\end{equation}
in a resummed form.  It is easy to confirm that the expansion of
Eq.~\eqref{eq:diff} up to the first order in $\Dth$ is precisely the
discretized version of the differential
equation~\eqref{eq:diff_eq_w}.  We use this resummed form to enhance
the numerical stability, which is improved by the fact that $\Dth$
appears in the denominator.

Our goal at the present is to integrate Eq.~\eqref{eq:diff}
numerically and make it sure that the resulting propagator is
non-vanishing for $\nu'=-\nu$ and is expected to be the analytical
solution~\eqref{eq:green} or its discretized representation:
\begin{equation}
 G(\nu,\theta) = \Nt\,\frac{\rmi}
  {\omega_{\rm min}^2 (\nu^2-\mu^2) + \rmi\epsilon} \Bigl[
  1-\rme^{2\rmi\omega_{\rm min}^2(\nu^2-\mu^2)\theta - 2\epsilon\theta} \Bigr]\;.
\label{eq:green_lat}
\end{equation}
We note that a dimensionless mass parameter $\mu$ is not necessarily
an integer, while $\nu$ is quantized corresponding to the Fourier mode
under the periodic boundary condition (see Eq.~\eqref{eq:munu} for
definition).

\begin{figure}
 \begin{center}
 \includegraphics[width=0.49\textwidth]{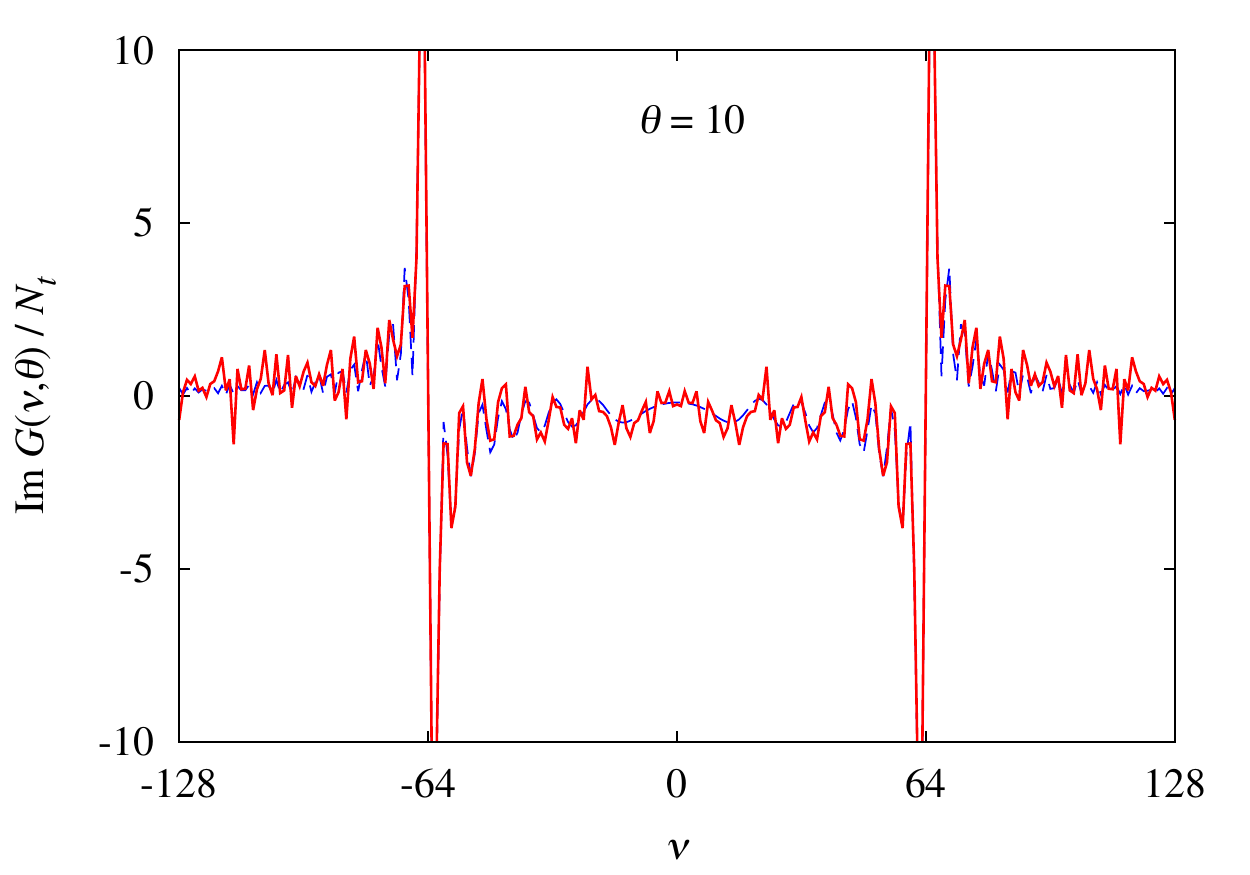}
 \includegraphics[width=0.49\textwidth]{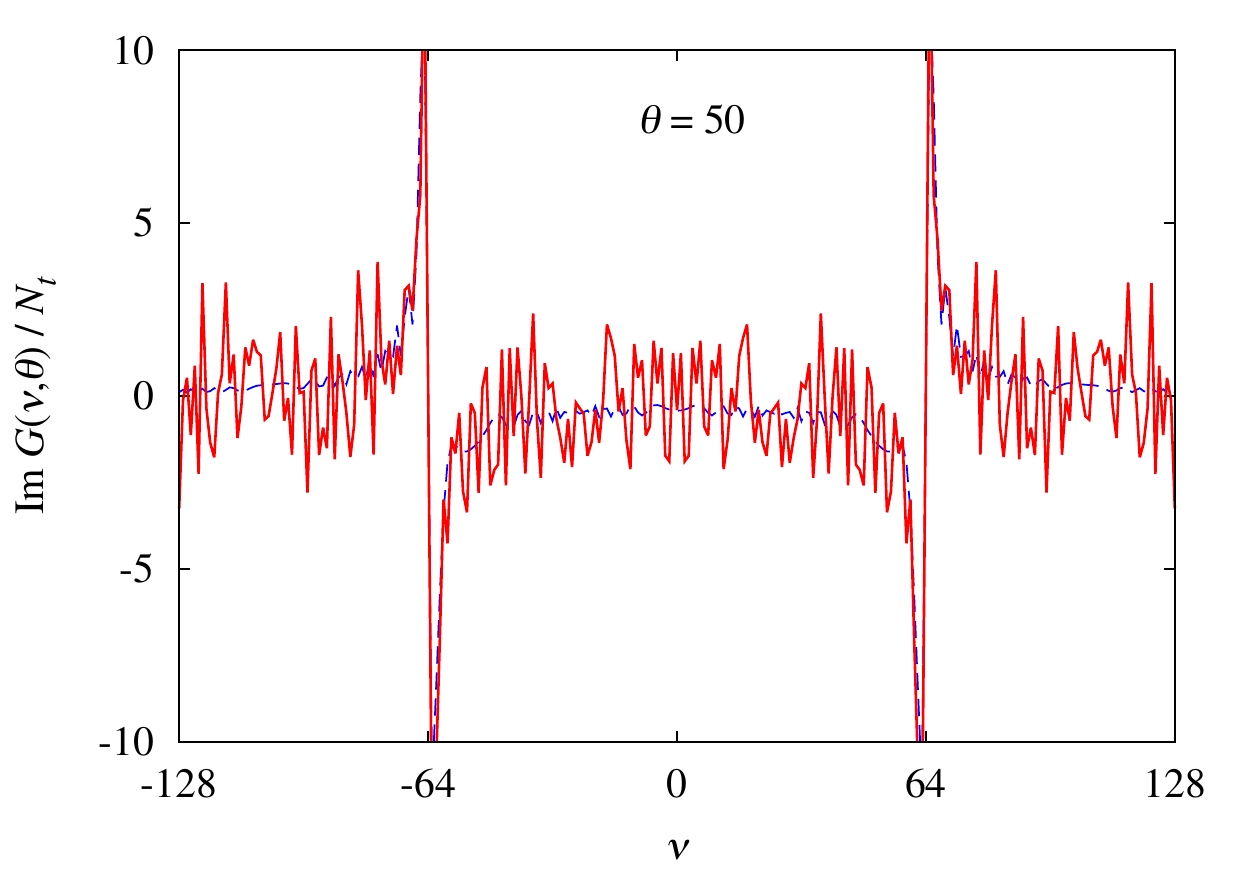}
 \end{center}
 \caption{Numerical solution for the free propagator with the
   ensemble average over 1000 independent runs with $\epsilon=10^{-2}$
   at $\theta=10$ (left) and $\theta=50$ (right).  We choose
   $\Dth=10^{-2}$ and $\mu=64$.}
 \label{fig:prop}
\vspace{5mm}
\end{figure}

Now let us consider the propagator for a specific choice of
parameters; $\mu=64$ and $\Nt=256$.  We show the imaginary part of
$G(\nu,\theta)$ from our numerical results in Fig.~\ref{fig:prop};  we
take the ensemble average over 1000 independent runs with
$\Dth=10^{-2}$ and $\epsilon=10^{-2}$.  We can see that the results at
$\theta=10$ (left of Fig.~\ref{fig:prop}) turns out to be quite
consistent with the analytical expectation from
Eq.~\eqref{eq:green_lat}.  We emphasize that our numerical results in
the left of Fig.~\ref{fig:prop} even reproduce the fine structure of
oscillation term, $\rme^{2\rmi\omega_{\rm min}^2\nu^2\theta}$, because
the damping factor $\rme^{-2\epsilon\theta}\simeq 0.82$ is not small.
Therefore we should take either a larger $\epsilon$ or a larger
$\theta$ for better convergence.  If $\epsilon$ gets larger, however,
the propagator poles would become obscure.  If we continue our
simulation till larger $\theta$, the numerical results suffer from
severe fluctuations as shown in the right of Fig.~\ref{fig:prop}
(which is an example at $\theta=50$).  These rough results are caused
not by numerical instability but merely by statistical problem.  As we
evolve the field value with increasing $\theta$, we accumulate all
contributions from $\eta_\nu(\theta)$ at each step of $\theta$.  This
means that we need to prepare more independent runs with increasing
$\theta/\Dth$ to get convergent results.  It is therefore a
time-consuming task to evolve the system up to $\theta=10^6$ to
suppress unwanted oscillations.

\begin{figure}
 \begin{center}
 \includegraphics[width=0.6\textwidth]{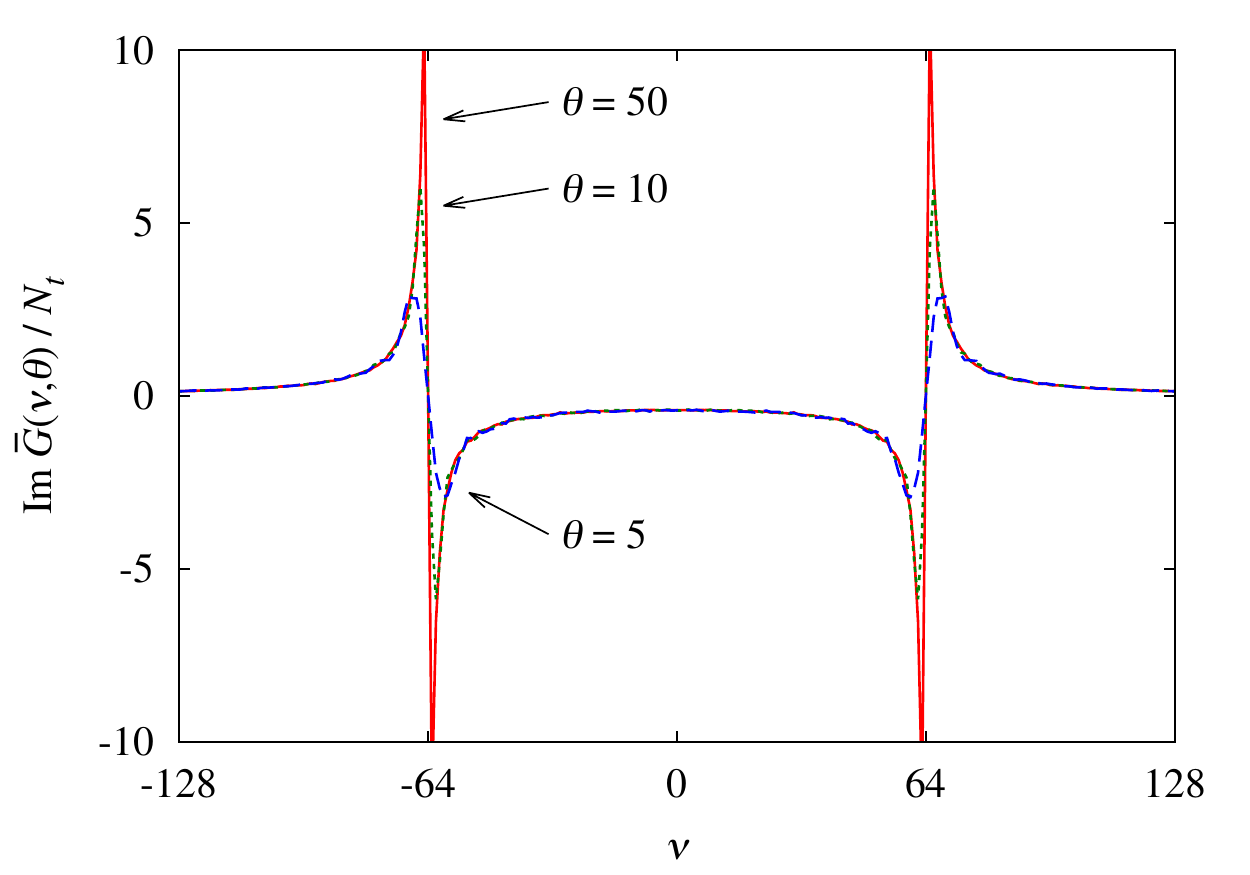}
 \end{center}
 \caption{Numerical results for the $\theta$-averaged propagator with
   the ensemble average still taken over 1000 independent runs.
   Dashed, dotted, and solid curves represent the results at
   $\theta=5$, $10$, and $50$, respectively.  We choose $\Dth=10^{-2}$
   and $\mu=64$ again.}
 \label{fig:propav}
\vspace{5mm}
\end{figure}

We can reduce the computational cost by averaging out to get rid of
the oscillatory part out from the vacuum contribution.  That is, we
see that the oscillatory part quickly disappears once we take a
$\theta$-average that is defined by
\begin{equation}
 \overline{\langle \calO[\phi(x,\theta)]
 \rangle_\eta} \equiv \frac{1}{\theta}\int_0^\theta \rmd\theta\,
  \langle\calO[\phi(x,\theta)] \rangle_\eta \;,
\label{eq:average}
\end{equation}
which is supposed to coincide with the correct expectation value for a
large value of $\theta$ if the unnecessary terms killed by a finite
$\epsilon$ are always accompanied by $\theta$-oscillation.  This is a
common practice and most of numerical works in the
literature~\cite{Berges:2005yt} makes use of this trick.
Nevertheless, strictly speaking, it is a non-trivial question whether
this procedure of taking the $\theta$-average is always harmless not
affecting the physical answer for any operators.  To the best of our
knowledge we do not have a general proof but we have performed
explicit calculations for the one-loop self-energy to confirm that
Eq.~\eqref{eq:average} gives the correct answer, which is explained in
details in Appendix~\ref{app:oneloop}.

The example in Fig.~\ref{fig:prop} evidently indicates the necessity
of taking the $\theta$-average to acquire converging results within
reasonable machine time.  Now we make use of the $\theta$-averaging
procedure of Eq.~\eqref{eq:average} to compute the propagator, which
is plotted in Fig.~\ref{fig:propav}.  It is obvious at a glance that
the simulation quickly converges to the smooth curve of the free
propagator already around $\theta\sim 10$.

\section{Inclusion of Interaction Effects}
\label{sec:interaction}

Let us continue our discussions with interaction effects using the
same quantum mechanical (0+1 dimensional) example where no ultraviolet
divergence appears and thus the theoretical setup is clean.  The
Langevin equation with $\lambda\neq 0$ reads in frequency space:
\begin{equation}
 \begin{split}
 \partial_\theta\,\phi_\nu(\theta) &= \rmi (\omega^2
  -\xi^2 + \rmi\epsilon)\,\phi_\nu(\theta) \\
  &\qquad -\frac{\rmi\lambda}{\Nt^2} \sum_{\nu_1,\nu_2}
  \phi_{\nu_1}(\theta)\, \phi_{\nu_2}(\theta)\,
  \phi_{\nu-\nu_1-\nu_2}(\theta) + \eta_\nu(\theta) \;.
 \end{split}
\label{eq:diff_eq_w_int}
\end{equation}
The question is how to discretize the above differential equation
avoiding numerical instability.  If we simply add the interaction term
on top of our procedure in the previous section, numerical instability
badly grows up for $\Dth=10^{-2}$  (but a smaller $\Dth$ like
$10^{-5}$ can stabilize the simulation).

\subsection{Exact and approximated results}

\begin{figure}
 \begin{center}
    \includegraphics[width=0.6\textwidth]{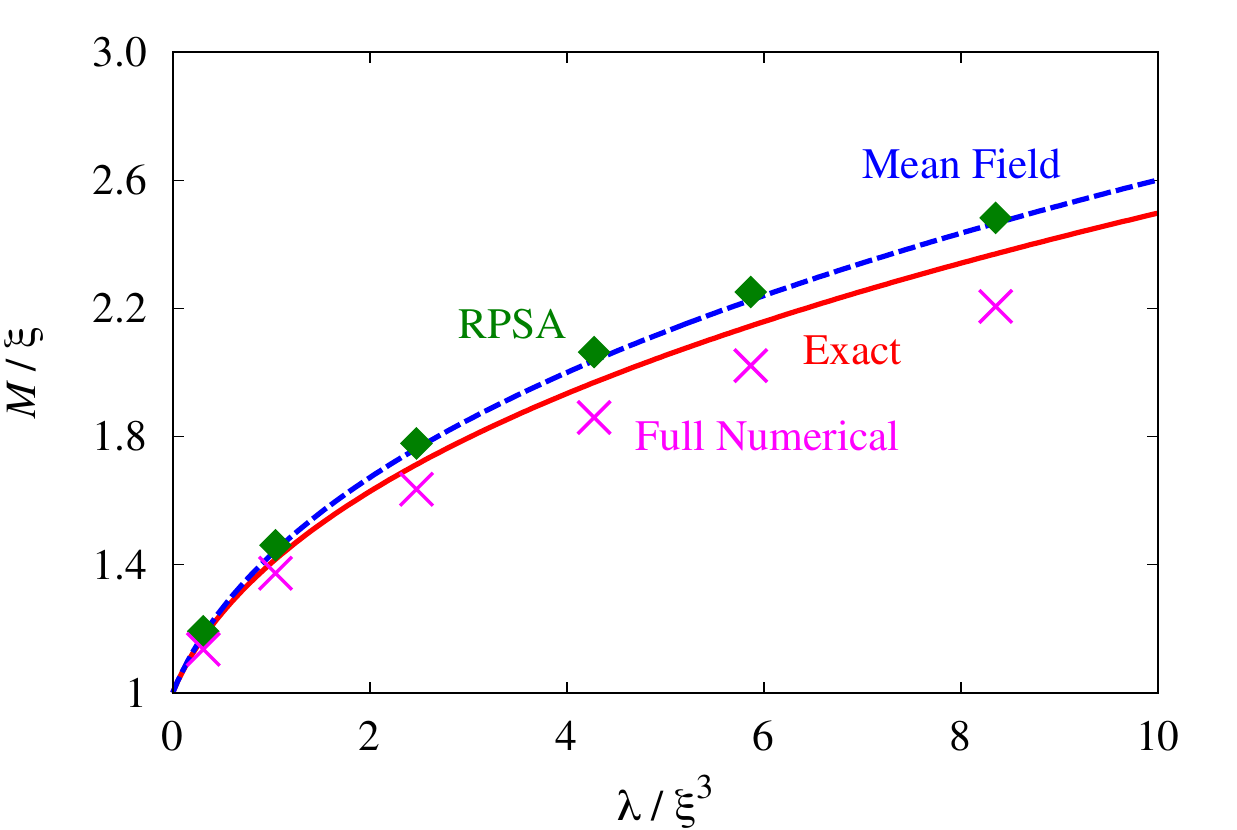}
 \end{center}
 \caption{Effective mass, $M$, as a function of the coupling $\lambda$
   in the unit of $\xi$.  The solid curve represents the exact answer
   by diagonalization of the Hamiltonian, the dashed one is the
   self-consistent solution~\eqref{eq:consistent}.  The filled
   diamonds represent the results from the restricted phase-space
   approximation we are proposing, while the crosses are the full
   numerical results without truncation;  see the text for more
   explanations.}
 \label{fig:mass}
\vspace{5mm}
\end{figure}

In this simple system we can find the ``exact'' answer by
diagonalizing the Hamiltonian using the harmonic oscillator bases,
which is elucidated in details in Appendix~\ref{app:full}.
Interestingly, in this case, the mean-field approximation or the
Hartree approximation would lead to results surprisingly close to the
exact answer.  In this approximation the interaction effects are
assumed to be all renormalized in the effective mass.  In the one-loop
level the self-energy in the continuum theory (which is already a good
approximation in our setup with $\Nt=256$) is found as
\begin{equation}
 \Pi = \rmi (-\rmi 6\lambda)\cdot\frac{1}{2} \int\frac{\rmd\omega}
  {2\pi}\; \frac{\rmi}{\omega^2 - \xi^2 + \rmi\epsilon}
 = \frac{3\lambda}{2\xi} \;.
\label{eq:onemass}
\end{equation}
We note that this one-dimensional integration results in a finite
number.  Hence, the effective mass should be shifted by $M^2=\xi^2+\Pi$
at the one-loop order.  In the mean-field resummation, the one-loop
tadpole diagrams are all taken into account through the
self-consistency condition or the gap equation,
\begin{equation}
 M^2 = \xi^2 + \frac{3\lambda}{2M} \;,
\end{equation}
in which the bare mass in Eq.~\eqref{eq:onemass} is replaced with the
effective mass $M$.  We can write the analytical solution of the gap
equation down as
\begin{equation}
 M =
  \biggl( \frac{3\lambda}{4} + \sqrt{\frac{9\lambda^2}{16}
  -\frac{\xi^6}{27}} \biggr)^{1/3}
  + \biggl( \frac{3\lambda}{4} - \sqrt{\frac{9\lambda^2}{16}
  -\frac{\xi^6}{27}} \biggr)^{1/3} \;.
\label{eq:consistent}
\end{equation}
We plot the dimensionless $M/\xi$ as a function of the dimensionless
coupling $\lambda/\xi^3$ in Fig.~\ref{fig:mass}.  From this we can
deduce how much the effective mass $M$ is enhanced from the bare mass
$\xi$.  For example, if we use $\lambda=0.5$ and
$\xi = \omega_{\rm min}\cdot\mu$ with $\mu=24$ and $64$, the
dimensionless coupling is $\lambda/\xi^3\simeq 2.48$ and $0.13$.
Then, multiplying the enhancement factor inferred from
Fig.~\ref{fig:mass}, we can get the mean-field masses as
$M = \omega_{\rm min}\cdot\mu'$ with $\mu'\approx 42.3$ and $69.5$,
respectively.  We will confirm these estimates soon later.

We here would like to draw an attention to the fact that the
mean-field results are amazingly close to the numerically exact
answer.  This nice agreement is attributed to the behavior of the full
numerical solution of this anharmonic oscillator problem;  the residue
of the propagator hardly deviates from the unity and the imaginary
part in the self-energy does not arise due to the phase space
limitation.  In other words, in the language of the 0+1 dimensional
field theory, the wave-function renormalization is negligibly small in
this particular case.

\subsection{Numerical results with the full interaction}
\label{sec:results_full}

We can take account of the self-interaction terms by adding them to
Eq.~\eqref{eq:diff} as they appear in Eq.~\eqref{eq:diff_eq_w_int}.
However, this straightforward implementation is not very stable for a
long time run.  We find that it would be much advantageous to add the
interaction terms in original $t$-space by taking the Fourier
transformation back.  The interaction is local then, while many
non-local terms are involved in $\omega$-space as in
Eq.~\eqref{eq:diff_eq_w_int}.

For concrete procedures of the updates, we first prepare
$\phi(t,\theta)$ and its Fourier transform $\phi_\nu(\theta)$.  Then
we calculate the difference from the kinetic term in $\omega$-space as
\begin{equation}
 \phi_\nu(\theta) + \delta\phi_\nu(\theta) = \frac{\rme^{-\epsilon\Dth}}
 {1-\rmi\omega_{\rm min}^2(\nu^2-\mu^2)\,\Dth}\,\phi_\nu(\theta) \;.
\end{equation}
Also we calculate the difference coming from the interaction terms in
$t$ space as
\begin{align}
 \phi(t,\theta) + \delta\phi(t,\theta) &= \phi(t,\theta) -\rmi\lambda
  \phi^3(t,\theta) \Dth + \eta(t,\theta)\,\Dth \notag\\
 & \approx \frac{\phi(t,\theta)}
  {\sqrt{1 + 2\rmi\lambda {\phi}^2(t,\theta)\Dth}}
  + \eta(t,\theta)\,\Dth \;.
\label{eq:resummed}
\end{align}
The numerical instability occurs when $\phi^3(t,\theta)\Dth$ happens
to take a large number.  The above resummed form is convenient to
avoid such a problem of instability.  We make a remark that this
special form solves
$\partial\phi/\partial\Dth = -\rmi\lambda\phi^3$.  Thanks to the
stability we can adopt $\Dth=10^{-2}$ below.

\begin{figure}
 \begin{center}
 \includegraphics[width=0.6\textwidth]{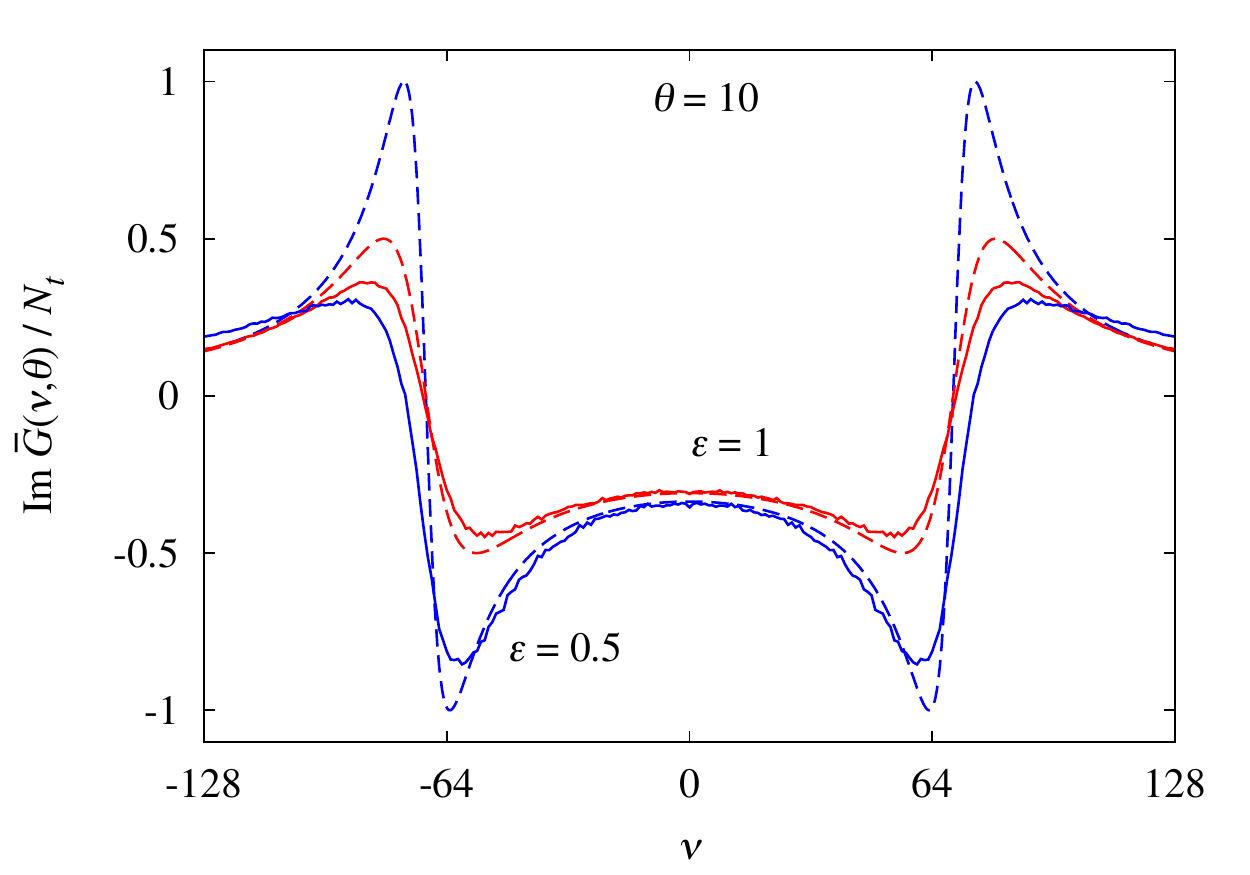}
 \end{center}
 \caption{Numerical results (by solid curves) for the
   $\theta$-averaged full propagators with $\mu=64$ for $\epsilon=0.5$
   and $1.0$ at $\theta=10$ with the ensemble average taken over 1000
   independent runs.  The interaction strength is chosen to be
   $\lambda=0.5$.  The dashed curves represent the (free) propagators
   with the effective mass $M=69.5\omega_{\min}$ (mean-field value)
   and corresponding $\epsilon$.}
 \label{fig:propfull}
\vspace{5mm}
\end{figure}

In this way we can calculate the propagator with full interaction
effects for $\lambda=0.5$ and various values of $\xi$ (or $\mu$).
Figure~\ref{fig:propfull} is an example of our simulation for
$\mu=64$.  Because there is no allowed phase space in spatial
zero-dimension, the physical width should be vanishing even in the
fully interacting case.  Our simulation results, however, exhibit some
unphysical width as seen in Fig.~\ref{fig:propfull}, while we can get
reasonable results if we force $\epsilon$ to be as large as $\sim 1$.
When $\epsilon$ becomes $0.5$ or smaller, the full results start
differing from the expected exact ones:  not only the unphysical width
appears but the propagator residue is also rotated with some complex
number, which can be fitted well by
\begin{equation}
 G(\omega) = \frac{\rmi A}{\omega^2 - M^2 + \rmi \Gamma} \;.
\label{eq:ansatz}
\end{equation}
As we described, as long as $\epsilon\gtrsim 1$, we find $A\sim 1$ and
$\Gamma\sim \epsilon$.  For smaller $\epsilon$ about $10^{-2}$, the
residue $A$ turns out to be $\sim -2\rmi$ and $\Gamma$ is of order
of the unity.  We have also reconfirmed this behavior using the upper
of Eq.~\eqref{eq:resummed} directly with $\Dth=10^{-5}$, and so we can
say that unphysical $A$ and $\Gamma$ for small $\epsilon$ are induced
not only under the resummation of Eq.~\eqref{eq:resummed}.  There
seems to be an unphysical fixed-point in the theory itself.

Although the shape of (the imaginary-part of) the propagator has funny
modifications with $A$ and $\Gamma$, the effective mass $M$ turns out
to be still close to the right value.  We have performed the fitting
between our numerical results and the ansatz of Eq.~\eqref{eq:ansatz}
for $\lambda=0.5$ with $\mu=48, 32, 24, 20, 18, 16$.  The global
behavior of $M$ obtained from the fit with the full numerical results
is fairly consistent with the exact answer as seen in
Fig.~\ref{fig:mass}.  We note that these are results for a choice of
$\epsilon=0.1$, but other values of $\epsilon$ would make only a tiny
quantitative difference.

\subsection{Unphysical fixed-point}
\label{sec:attractor}

The transparent method to figure the flow pattern out with changing
$\theta$ is to find the fixed-points using the Fokker-Planck
equation~\cite{Aarts:2013uza} which describes the equivalent dynamics
as the Langevin equation does.  One can prove that the
probability function, $P[\phir,\phii]$, should obey the following
differential equation (see Ref.~\cite{Damgaard:1987rr} for derivations
and also references therein for further details):
\begin{align}
 & \frac{\partial P}{\partial\theta} = \int\frac{\rmd\omega}{2\pi}
  \biggl[ \frac{\delta}{\delta\phir(\omega)} \biggl(\mathrm{Im}
  \frac{\delta S}{\delta \phi(-\omega)}
  + \frac{\delta}{\delta \phir(-\omega)} \biggr)
  - \frac{\delta}{\delta \phii(\omega)}\, \mathrm{Re}
  \frac{\delta S}{\delta \phi(-\omega)} \biggr] P \notag\\
 &= \int\frac{\rmd\omega}{2\pi} \biggl\{ 2\pi\delta(0)\cdot 2\epsilon
  -12\lambda \int\frac{\rmd\omega'}{2\pi} \phir'\phii'
  + \frac{\delta^2}{\delta\phir^2}
  + \epsilon\biggl(\phir\frac{\delta}{\delta\phir}
  + \phii\frac{\delta}{\delta\phii}\biggr) \notag\\
 &\quad + (\omega^2-m^2)\biggl(\phii\frac{\delta}{\delta\phir}
  -\phir\frac{\delta}{\delta\phii} \biggr)
  -\lambda\int\frac{\rmd\omega'\rmd\omega''}{(2\pi)^2}
  \bigl( 3\phir'''\phir'\phii'' - \phii'''\phii'\phii'' \bigr)
  \frac{\delta}{\delta\phir} \notag\\
 &\qquad\quad -\lambda\int\frac{\rmd^2\omega'\omega''}{(2\pi)^2}
  \bigl( 3\phir'''\phii'\phii'' - \phir'''\phir'\phir'' \bigr)
  \frac{\delta}{\delta\phii} \biggr\} P \;,
\label{eq:fokker-planck}
\end{align}
where we shortened our notation by writing $\phir'$ and $\phii'$ to
denote fields with the frequency $\omega'$, $\phir''$ and $\phii''$
with $\omega''$, and $\phir'''$ and $\phii'''$ with
$\omega-\omega'-\omega''$, with an exception that $\phir'\phii'$ in
the first line represents $\phir(-\omega')\phii(\omega')$.  In the end
of the $\theta\to\infty$ limit, $P[\phir,\phii;\theta]$ should
converge to an asymptotic form at the fixed-point.  In the free case
with $\lambda=0$ it is easy to confirm that the following probability
function:
\begin{equation}
 P[\phir,\phii] = N\exp\left[ -\epsilon\int\frac{\rmd\omega}{2\pi}
  (\phir,\phii) \begin{pmatrix}
  1 & -\frac{\epsilon}{\omega^2-m^2} \\
  -\frac{\epsilon}{\omega^2-m^2} &
  ~~1+\frac{2\epsilon^2}{(\omega^2-m^2)^2} \end{pmatrix}
  \begin{pmatrix} \phir \\ \phii \end{pmatrix} \right]
\label{eq:free_P}
\end{equation}
solves $\partial P/\partial\theta=0$.  Because we have numerically
found in the previous subsection that our results support a fitting
ansatz of Eq.~\eqref{eq:ansatz}, we should be able to perform
fixed-point analysis in the parameter space spanned by $A$, $M$, and
$\Gamma$.  In this way, from $\partial P/\partial\theta=0$, we can
derive equations for these ``variational parameters'' that play the
role of the gap equations.  The full analysis has turned out to be
quite complicated and we would like to leave detailed descriptions for
a separate paper. 

The Fokker-Planck equation~\eqref{eq:fokker-planck} is, however,
sufficiently useful for us to understand why our numerical results in
Sec.~\ref{sec:results_full} tend to fall into a wrong branch of $A$
and $\Gamma$.  In the free case we immediately see from
Eqs.~\eqref{eq:fokker-planck} and \eqref{eq:free_P} that the first
term involving $2\epsilon$ is exactly canceled by
$\delta^2/\delta\phir^2$ hitting on the exponential part of
Eq.~\eqref{eq:free_P}.  Once we modify $P[\phir,\phii]$ to allow for a
complex residue $A=|A|\rme^{\rmi\alpha}$ and a width $\Gamma$, the
corresponding probability function has a $\phir^2$-component with a
mixture of $\Gamma\cos\alpha$ and $(\omega^2-m^2)\sin\alpha$.  Thus,
$\delta^2/\delta\phir^2$ would generate a very singular term
proportional to $\int\rmd\omega\omega^2\sin\alpha$, so that
$\partial P/\partial\theta=0$ is easily achieved by an appropriate
(but unphysical) choice of $\alpha\neq0$ because its coefficient
$\int\rmd\omega\omega^2$ is overwhelming.  This is an intuitive
explanation of how an unphysical fixed-point cannot be avoidable with
a complex residue of the propagator, as we numerically observed in the
previous subsection.

As long as such an unphysical fixed-point is well separated from the
physical trajectories, the numerical simulation can correctly identify
the physical fixed-point, which is the case in the free theory.  The
flow structure around the fixed-points or attractors becomes highly
involved for general $\lambda\neq0$, and when the physical region is
contaminated by the unphysical fixed-point, the simulation is stuck
with pathological behavior (a part of which might be cured by the
change of variables as discussed in Ref.~\cite{Aarts:2012ft}).  A more
complete study will be reported elsewhere (including the explicit
check of the locality and the Dyson equations; see
Refs.~\cite{Aarts:2009uq,Aarts:2011ax}).

\subsection{Restricted phase-space approximation (RPSA)}
\label{sec:restriction}

\begin{figure}
 \includegraphics[width=\textwidth]{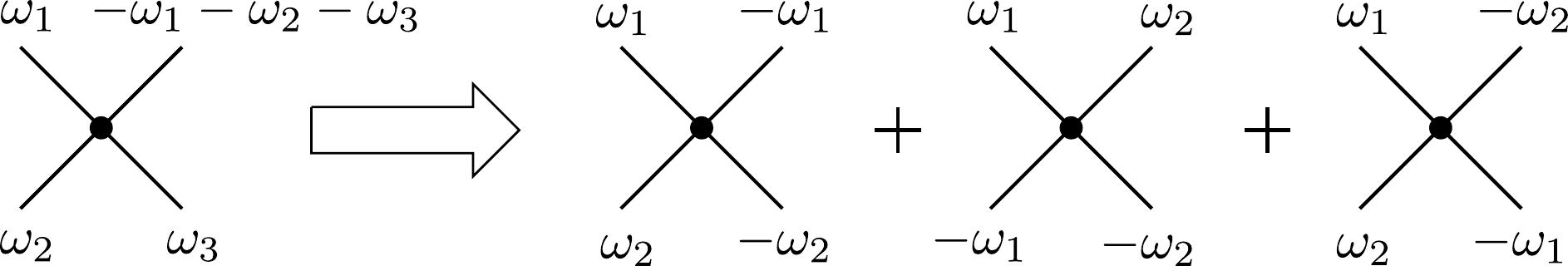}
 \caption{Schematic illustration for the RPSA that we propose in this
   paper.  The phase space associated with scattering is restricted to
   the limit of small angle (zero momentum transfer) in the $s$-,
   $t$-, and $u$-channels.  This treatment does not damage the
   essential features of real-time dynamics and even becomes exact in
   special models such as the large $N$ limit of $\mathrm{O}(N)$
   scalar model.}
\vspace{1em}
\label{fig:rpsa}
\end{figure}

Here we would propose a ``restricted phase-space approximation''
(RPSA) that is defined by the following truncation in the interaction
terms:
\begin{equation}
 \sum_{\nu_1,\nu_2}\phi_{\nu_1}(\theta)\,\phi_{\nu_2}(\theta)\,
  \phi_{\nu-\nu_1-\nu_2}(\theta)
 = 3\sum_{\nu_1} \phi_{-\nu_1}(\theta)\,\phi_{\nu_1}(\theta)\,
  \phi_{\nu}(\theta) + \text{(others)} \;.
\end{equation}
In the RPSA we discard terms referred to as ``others'' in the above.
We emphasize that this truncation should not damage the essential
features of real-time dynamics; in fact, if we work in the
$\mathrm{O}(N)$ scalar theory and take the limit of $N\to\infty$, only
the daisy diagrams remain in the leading order of the $1/N$ expansion
and discarded terms are all dropped off.  Therefore, the RPSA becomes
exact in this special case.  We present Fig.~\ref{fig:rpsa} for a
schematic illustration of the RPSA for the $\phi^4$ interaction.

In this prescription of the RPSA we can express the differential
equation as if it were a free-theory problem with a renormalized
mass-like term; i.e.,
\begin{align}
 \phi_\nu(\theta+\Dth) &= \frac{\rme^{-\epsilon\Dth}}
  {1-\rme^{\rmi(\omega^2-{\bar\xi}^2)}}\,\phi_\nu(\theta)
  + \eta_\nu(\theta) \;,\\
 {\bar\xi}^2 &\equiv \xi^2 + \frac{3\lambda}{\Nt^2} \sum_{\nu_1}
  \phi_{-\nu_1}(\theta)\,\phi_{\nu_1}(\theta) \;.
\end{align}
It should be mentioned that ${\bar\xi}^2$ is not a mass but it still
involves interactions.  So, the RPSA is not a mean-field approximation
and ${\bar\xi}$ is not a mean-field mass.  The point is that we can
treat ${\bar\xi}$ in the same way as the mass in the numerical
procedure.  This implies that the numerical simulation is stable even
with non-zero $\lambda$ as long as it is stable for a free theory.

\begin{figure}
 \includegraphics[width=0.49\textwidth]{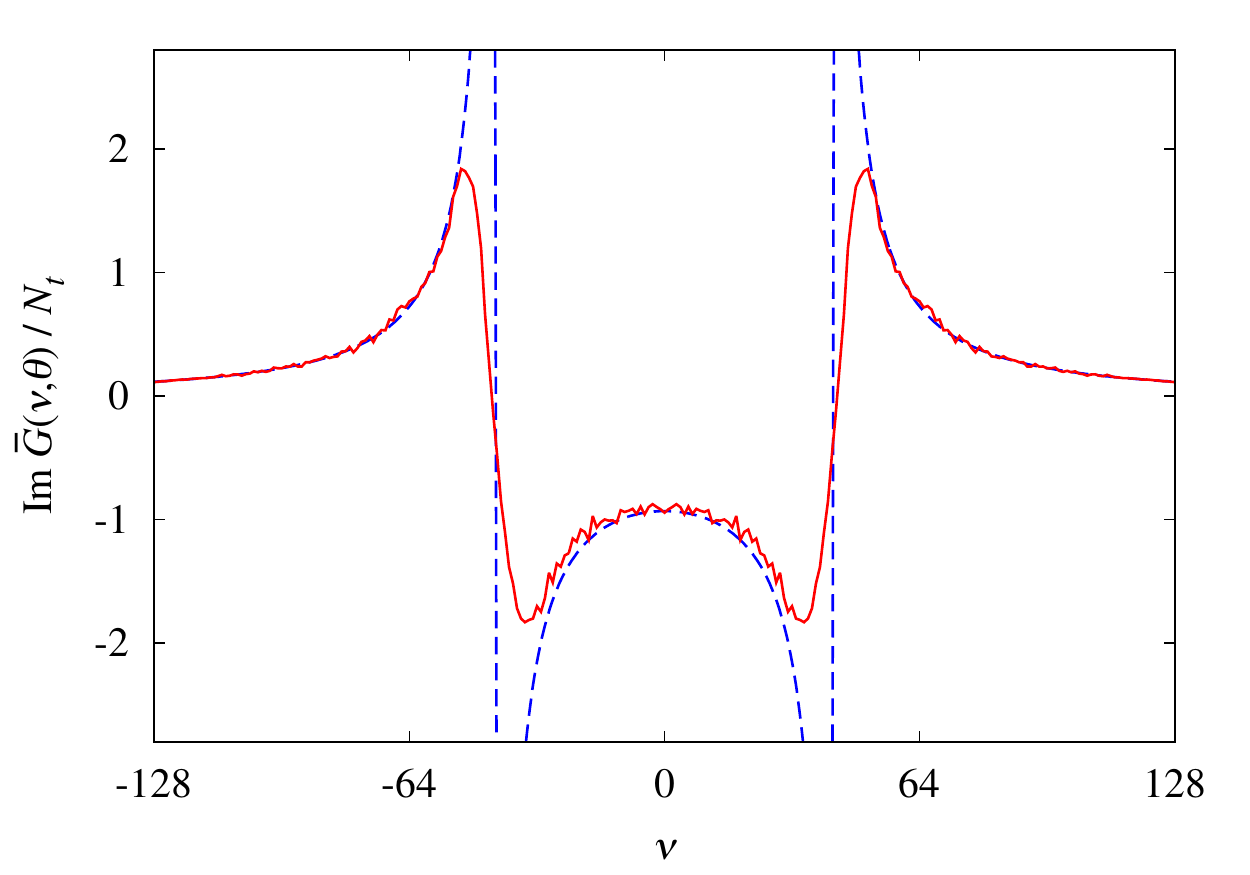}
 \includegraphics[width=0.49\textwidth]{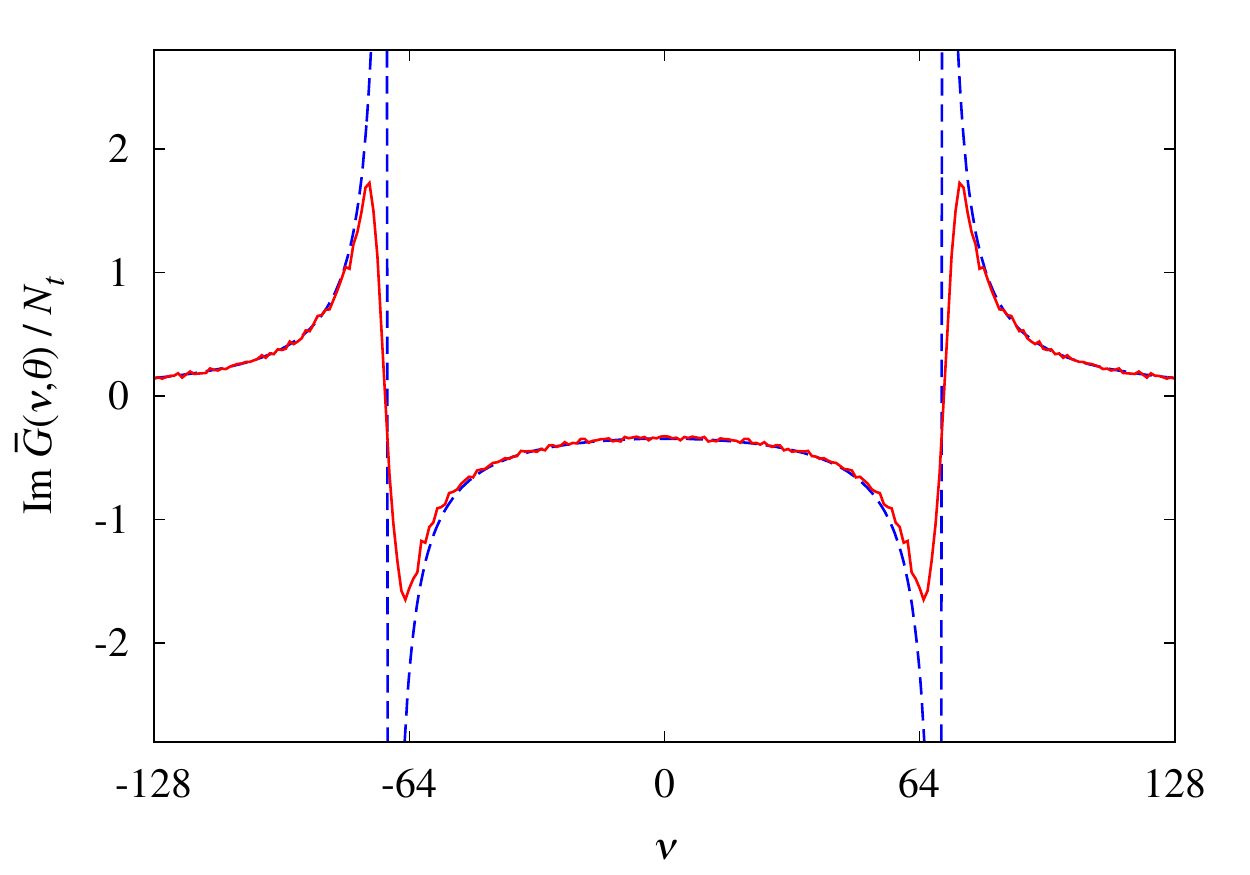}
 \caption{Numerical results (by solid curves) for the
   $\theta$-averaged propagator at $\theta=5$ with the ensemble
   average taken over 1000 independent runs.  We adopt $\xi=24$ (left)
   and $\xi=64$ (right), respectively.  The interaction strength is
   chosen to be $\lambda=0.5$.  The dashed curves represent the
   mean-field propagator.}
\label{fig:prop_hartree}
\end{figure}

Figure~\ref{fig:prop_hartree} shows our numerical results in the RPSA
for the propagator with interaction $\lambda=0.5$ and the bare mass
$\mu=24$ (left) and $\mu=64$ (right), respectively.  In view of these
results we can make sure that unphysical width is suppressed, and
indeed, the peak becomes sharper if we extend the simulation till a
further larger value of $\theta$.  So, we here manage to avoid the
unphysical fixed-point.

It is quite impressive that our numerical results agree well with the
mean-field propagator in which the mean-field masses are plugged;
namely, $\mu'\approx 42.3$ for $\mu=24$ and $\mu'=69.5$ for $\mu=64$
as shown by dashed curves in Fig.~\ref{fig:prop_hartree}.

Let us make this kind of comparison more quantitative.  We can fit the
numerical results using the parametrization of Eq.~\eqref{eq:ansatz}.
In this case of RPSA we find that $A$ is always close to the unity and
$\Gamma$ is as small as $\epsilon$ once we continue the simulation up
to a sufficiently large $\theta$.  Then, we deduce the effective
masses $M$ corresponding to $\mu=48, 32, 24, 20, 18, 16$ and put
crosses on Fig.~\ref{fig:mass}.  Surprisingly, the resultant $M$ turns
out to be on top of the mean-field prediction, though the RPSA is not
really equivalent to the mean-field approximation.



\section{Closed-time Path Formalism}
\label{sec:closed}

Now we have understood how we can correctly reproduce the free
propagator in frequency space with help of the $\rmi\epsilon$
prescription.  In this section let us briefly sketch how we can apply
the closed-time path formalism to compute the vacuum expectation value
directly in a form of Eq.~\eqref{eq:ex_O} not relying on the
$\rmi\epsilon$ prescription, which would be extended to more general
non-equilibrium environments.

For simplicity we limit ourselves only to the free theory because we
can carry out analytical evaluations using an explicit form of the
vacuum wave-function:
\begin{equation}
 \Psi_0[\phi] \equiv
  \langle\phi|\Omega\rangle = \Bigl(\frac{m}{\pi}\Bigr)^{1/4}
   \exp\Bigl(-\frac{m}{2}\,\phi^2 \Bigr) \;.
\label{eq:gaussian}
\end{equation}
We shall see the recovery of the free propagator in what follows.  We
compute the two-point function with inserting the complete sets;
$\int\rmd\phi_{\rm i}|\phi_{\rm i}\rangle\langle\phi_{\rm i}|$ and
$\int\rmd\phi_{\rm i}'|\phi_{\rm i}'\rangle\langle\phi_{\rm i}'|$ as
\begin{equation}
 \langle\Omega| \phi(\tf)\,\phi(\ti)|\Omega\rangle
  = \sqrt{\frac{m}{\pi}}\int \rmd\phi_{\rm i}\,\rmd\phi'_{\rm i}\,
   \rme^{-m(\phi^{\prime 2}_{\rm i}+\phi^2_{\rm i})/2} \phi_{\rm i}
  \langle\phi'_{\rm i}| \rme^{-\rmi H(\ti-\tf)} \phi\, \rme^{-\rmi H(\tf-\ti)}
  |\phi_{\rm i}\rangle \;.
\label{eq:element}
\end{equation}
Then, using the Fock space bases, we can write the matrix element
appearing in the above expression as
\begin{align}
 & \sum_{n,n'}\langle\phi'_{\rm i}| n'\rangle\langle n'|
  \rme^{-\rmi H(\ti-\tf)} \phi\, \rme^{-\rmi H(\tf-\ti)}
  |n\rangle\langle n| \phi_{\rm i}\rangle \notag\\
 &= \rme^{-\rmi m (\tf-\ti)} \sum_n
  \langle\phi'_{\rm i}| n\!-\!1\rangle\langle n\!-\!1|\phi|n\rangle
  \langle n|\phi_{\rm i}\rangle
   +\rme^{\rmi m(\tf-\ti)} \sum_n
  \langle \phi'_{\rm i}|n\!+\!1\rangle\langle n\!+\!1|\phi|n\rangle
  \langle n|\phi_{\rm i}\rangle \notag\\
 &
  = \cos[m(\tf-\ti)]\, \phi'_{\rm i}\,
  \delta(\phi_{\rm i}-\phi'_{\rm i})
   -\rmi \sin[m(\tf-\ti)]\,\frac{\rmd}{m\, \rmd\phi'_{\rm i}}
    \delta(\phi_{\rm i}-\phi'_{\rm i}) \;.
\label{eq:matrix_ele}
\end{align}
We can readily reach the final answer as follows:
\begin{equation}
 \langle\Omega|\phi(\tf)\,\phi(\ti)|\Omega\rangle
  = \rme^{-\rmi m(\tf-\ti)}\,\sqrt{\frac{m}{\pi}}\int \rmd\phi_{\rm i}\,
   \phi^2_{\rm i}\rme^{-m\phi^2_{\rm i}}
  = \frac{1}{2m}\,\rme^{-\rmi m(\tf-\ti)} \;.
\label{eq:free_answer}
\end{equation}
This is an almost trivial example; nevertheless, it is far from
trivial to understand how the CTP formalism in stochastic quantization
works out to realize this simple propagator.  In the CTP formalism
the matrix element appearing in Eq.~\eqref{eq:element} is put into a
form of the functional integral (or the path integral in our
one-dimensional example) that is written as
\begin{align}
 \langle\phi'_{\rm i}| \rme^{-\rmi H(\ti-\tf)}\phi\,
  \rme^{-\rmi H(\tf-\ti)}|\phi_{\rm i}\rangle
 &= \int \rmd\phi_{\rm f}\,\rmd\phi'_{\rm f}
   \int_{\phi'_{\rm f}\to\phi'_{\rm i}} \calD\phi\,\rme^{\rmi S}\,
   \langle\phi'_{\rm f}|\phi|\phi_{\rm f}\rangle
   \int_{\phi_{\rm i}\to\phi_{\rm f}} \calD\phi\,\rme^{\rmi S}
   \notag\\
 &= \int\rmd\phi_{\rm f}\, \phi_{\rm f}
  \oint_{\phi_{\rm i}\to\phi_{\rm f}\to\phi'_{\rm i}} \calD\phi\,
  \rme^{\rmi S} \;.
\end{align}
Here $\int_{\phi_{\rm i}\to\phi_{\rm f}}$ represents the path integral
with a boundary with the initial condition $\phi=\phi_{\rm i}$ at
$t=\ti$ and the final condition $\phi=\phi_{\rm f}$ at $t=\tf$, etc. 
The standard knowledge on the path integral representation of quantum
mechanics tells us that this path integral part is nothing but the
Feynman kernel whose explicit form is
\begin{equation}
 \begin{split}
  & \int_{\phi_{\rm i}\to\phi_{\rm f}}\calD\phi\,\rme^{\rmi S}
   = \sqrt{\frac{m}{2\pi \rmi \sin[m(\tf\!-\!\ti)]}} \\
  &\qquad\times \exp\biggl\{ \frac{\rmi m}{2\sin[m(\tf-\ti)]}\Bigl[
    (\phi_{\rm i}^2+\phi_{\rm f}^2)\cos[m(\tf-\ti)]-2\phi_{\rm i}
    \phi_{\rm f} \Bigr]\biggr\} \;,
 \end{split}
\label{eq:Feynman}
\end{equation}
for the harmonic oscillatory (see Ref.~\cite{feynman1965quantum} for a
well-known textbook).  We can rederive the matrix
element~\eqref{eq:matrix_ele} using the above Feynman
kernel~\eqref{eq:Feynman}.  It is a bit tedious calculation, but quite
instructive, so we summarize the key equations of the derivation
below.

From the Feynman kernel~\eqref{eq:Feynman} the path integral along the
closed contour is found to take:
\begin{equation}
 \begin{split}
 & \oint_{\phi_{\rm i}\to\phi_{\rm f}\to\phi_{\rm i}}\calD\phi\,
  \rme^{\rmi S}
  = \frac{m}{2\pi\sin[m(\tf-\ti)]} \\
 &\quad\times \exp\biggl\{ \frac{\rmi m}{2\sin[m(\tf-\ti)]}
  \Bigl[ (\phi_{\rm i}^2-\phi_{\rm i}^{\prime 2})\cos[m(\tf-\ti)]
  -2(\phi_{\rm i}-\phi'_{\rm i})\phi_{\rm f} \Bigr]\biggr\} \;.
 \end{split}
\end{equation}
Then, we can perform the integration with respect to $\phi_{\rm f}$
that picks up the following part from the whole expression:
\begin{equation}
 \int\rmd\phi_{\rm f}\,\phi_{\rm f}\,\exp\biggl\{\frac{-\rmi m
   (\phi_{\rm i}-\phi'_{\rm i})}{\sin[m(\tf-\ti)]}\,\phi_{\rm f}
  \biggr\}
  = \frac{\sin^2[m(\tf-\ti)]}{\rmi m^2} \frac{\rmd}{\rmd\phi'_{\rm i}}
    2\pi\delta(\phi_{\rm i}-\phi'_{\rm i}) \;.
\end{equation}
Because $\phi'_{\rm i}$ is an integration variable in the convolution
with the initial wave function, we can move the derivative using the
integration by part to reach:
\begin{align}
 & \int\rmd\phi_{\rm f}\,\phi_{\rm f}\,
  \mathrm{Re}\oint_{\phi_{\rm i}\to\phi_{\rm f}\to\phi'_{\rm i}} \calD\phi\,
  \rme^{\rmi S}
  = \frac{m}{2\pi\sin[m(\tf-\ti)]} \notag\\
 &\quad \times \frac{\rmi m}{\sin[m(\tf-\ti)]}\cos[m(\tf-\ti)]
  \,\phi'_{\rm i} \cdot
  \frac{\sin^2[m(\tf-\ti)]}{\rmi m^2}\,2\pi\delta(\phi_{\rm i}
  -\phi'_{\rm i}) \notag\\
 &= \cos[m(\tf-\ti)]\,\phi_{\rm i}\,\delta(\phi_{\rm i}-\phi'_{\rm i}) \;.
\end{align}
The $\phi'_{\rm i}$ derivative acts also on the wave function, so that
it yields the imaginary part of Eq.~\eqref{eq:matrix_ele} in the same
way.  Then, we can explicitly see that we surely reproduce the matrix
element of Eq.~\eqref{eq:matrix_ele} and thus
Eq.~\eqref{eq:free_answer} as well.

This is how the CTP formalism works analytically to describe the time
evolution.  We focused on the vacuum expectation value in a free
theory, but the generalization is not difficult.  Even for more
complicated operators with interaction turned on, one can understand
that the most fundamental building block for the formalism is still
the matrix element~\eqref{eq:matrix_ele} of the free propagation;
therefore, we will concentrate on this quantity in our numerical
analysis below.

Just for the test purpose of the CTP formalism, we shall fix
$\phi_{\rm i}=\phi'_{\rm i}$ in the beginning and then calculate the
expectation value of $\phi(\tf)$ to pick up the diagonal component of
Eq.~\eqref{eq:matrix_ele}.  It should be mentioned that in stochastic
quantization we cannot directly calculate the amplitude such as the
Feynman kernel but it is always an ``expectation value'' of some
operator that we can estimate.  Thus, with a given boundary condition,
$\phi_{\rm i}=\phi'_{\rm i}$, if we compute the $\eta$-average of
$\phi(\tf)$, it should be interpreted as
\begin{equation}
 \frac{\langle\phi_{\rm i}| \rme^{-\rmi H(\ti-\tf)}\,\phi\,
  \rme^{-\rmi H(\tf-\ti)} |\phi_{\rm i}\rangle}
  {\langle\phi_{\rm i}|\phi_{\rm i}\rangle} = \phi_{\rm i}
  \cos[m(\tf-\ti)] \;.
\label{eq:costfti}
\end{equation}
We note that $\delta(0)$ cancels in this ratio.  This is the quantity
that we wish to reproduce in the following numerical test of
stochastic quantization.  That is, we will compute the left-hand side
of Eq.~\eqref{eq:costfti} to check if the numerical results coincide
with the right-hand side.  One might have a feeling that such a
numerical calculation only to have $\cos[m(\tf-\ti)]$ should have no
problem.  From the point of view of practical numerical procedures,
however, it is not really so because the time evolution emerges in a
finite extent of time between $\ti$ and $\tf$, and so the boundary
condition at $\tf$ is also necessary for the numerical derivative
there.  The CTP formalism provides us with a natural solution as we
will see soon later.

We discretize in $t$-space and replace the derivatives with
appropriate finite differences.  As is well known in the numerical
analysis of the diffusion equation, a na\"{i}ve replacement called the
Euler method is not stable depending on the ratio of $\Dth$ and
$\Dt$~\cite{Press:2007:NRE:1403886}.  It is a textbook knowledge how
to improve the numerical stability;  in the implicit method a part of
$\phi(t,\theta)$ and $\phi(t\pm\Dt,\theta)$ are replaced with
$\phi(t,\theta+\Dth)$ and $\phi(t\pm\Dt,\theta+\Dth)$, which
significantly enhances the stability.  Here, let us adopt a simple
algorithm; i.e., the half implicit method aka Crank-Nicolson method,
which is easier in our case than another common choice; that is, the
adaptive stepsize method~\cite{Aarts:2009dg}.  In this half implicit
method the Laplacian is concretely implemented via
\begin{equation}
 \sfD_{2\Nt+1} \!\!
 \begin{pmatrix}
  \phi(\ti,\theta\!+\!\Dth) \\
  \phi(\ti\!+\!\Dt,\theta\!+\!\Dth) \\
  \vdots \\
  \phi(\tf,\theta\!+\!\Dth)
 \end{pmatrix}
 = \bigl[ \sfD^\ast_{2\Nt+1} \!- \rmi\Dth \xi^2 \unity \bigr] \!
 \begin{pmatrix}
  \phi(\ti,\theta) \\
  \phi(\ti\!+\!\Dt,\theta) \\
  \vdots \\
  \phi(\tf,\theta)
 \end{pmatrix}
 + \Dth \!
\begin{pmatrix}
  \eta(\ti,\theta) \\
  \eta(\ti\!+\!\Dt,\theta) \\
  \vdots \\
  \eta(\tf,\theta)
 \end{pmatrix} .
\label{eq:diff_matrix}
\end{equation}
Hereafter, we rescale all variables such as $\xi$ to make them
dimensionless by multiplying a proper power of $\Dt$; i.e., we measure
all quantities in the unit of $\Dt$.  Here $\sfD_{2\Nt+1}$ represents
the derivative (Laplacian) matrix.  Including the forward and the
backward paths between $\ti$ and $\tf$ discretized with $\Nt$ sites,
$\sfD_{2\Nt+1}$ is a $(2\Nt+1)\times(2\Nt+1)$ matrix on the closed
path. Taking account of the change of the sign of $\Dt$ in the
backward direction, we can write the discretized matrix down as
\begin{equation}
 \sfD_{2\Nt+1} =
{\setlength{\arraycolsep}{6pt}
 \left( \begin{array}{cc|c|cc}
  \lsymb{\sfD_{\Nt}}&            &            &           &\hsymb{0}\\
                     &            &-\half\alpha&           &         \\
  \hline
                     &-\half\alpha& 1          &\half\alpha&         \\
  \hline
                     &            & \half\alpha&           &         \\
 \hsymb{0}           &            &            & \multicolumn{2}{r}{\lsymb{\sfD^\ast_{\Nt}}\quad}
 \end{array} \right)
} \;,
\label{eq:Ddb}
\end{equation}
where $\sfD_{\Nt}$ represents the $\Nt\times\Nt$ sub-matrix defined as
\begin{equation}
 \sfD_{\Nt} =
{\setlength{\arraycolsep}{6pt}
 \begin{pmatrix}
  1+\alpha     & -\half\alpha &               & \hsymb{0} \\
  -\half\alpha & 1+\alpha     & -\half\alpha  & \\
               & \ddots       & \ddots        & \ddots \\
  \hsymb{0}    &              & -\half\alpha  & 1+\alpha
 \end{pmatrix}
} \;.
\label{eq:D}
\end{equation}
We note that $\alpha$ is a pure-imaginary number given by
$\alpha \equiv -\rmi \Dth/\Dt^2$.  What we need is the field value at
the next step, $\theta+\Dth$, and thus, we can solve them by applying
$\sfD^{-1}_{2\Nt+1}$ on the both sides of Eq.~\eqref{eq:diff_matrix}.

\begin{figure}
\begin{center}
 \includegraphics[width=0.6\textwidth]{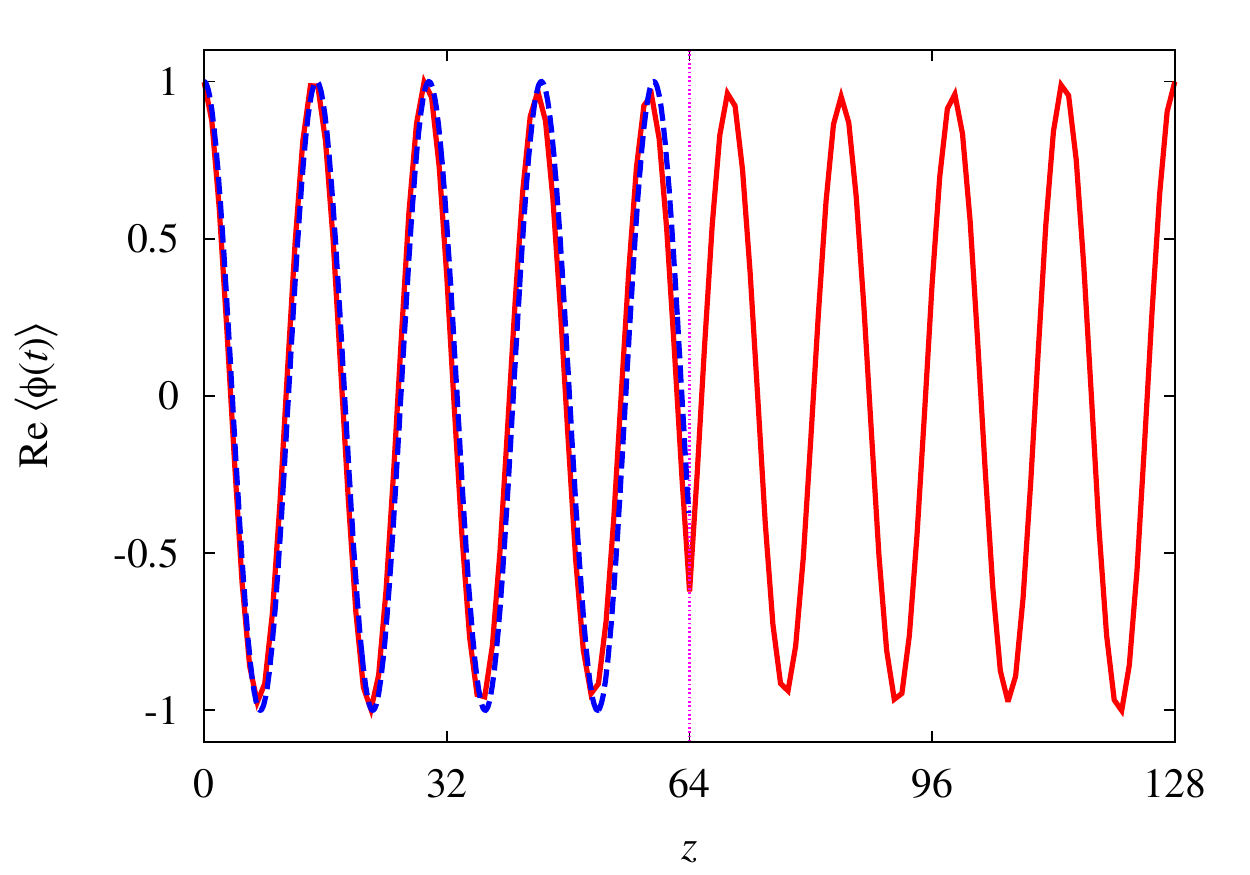}
\end{center}
 \caption{Simple demonstration of the CTP formalism;  the field
   profile at $\theta=500$.  The closed path is unfolded with $t$
   replaced with $z$ (see Fig.~\ref{fig:schematic2}) to separate the
   forward path ($t=\ti\to\tf$) and the backward path ($t=\tf\to\ti)$
   where $\tf=64$.  The ensemble average is taken over 100 runs.  The
   dashed curve represents $\phi_{\rm i}\cos(mt)$.}
\label{fig:ctp}
\vspace{5mm}
\end{figure}

We present our numerical results in Fig.~\ref{fig:ctp} in which the
time axis is unfolded from $t$ to $z$;  we should interprete $z>\tf$
as $t=z-\tf$ on the backward path returning to $\ti$.  For the results
in Fig.~\ref{fig:ctp} we choose $\Nt=64$ and so $z$ runs from $0$ to
$128$.  The initial value is chosen to be
$\phi_{\rm i}=\phi'_{\rm i}=1$.  The oscillation period is determined
by the mass parameter $\xi$ that is now fixed to be $\xi=4.25$, which
means that $4.25$ periods should appear between $t=0$ and $63$ as is
indeed the case in Fig.~\ref{fig:ctp}.  Because the
matrix~\eqref{eq:Ddb} has a special point at $t=\tf$, the derivative
jumps there, so that the time evolution is reflected from the forward
to the backward direction correctly.  This is in fact the reason why
the backward path is absolutely needed in the CTP formalism.  Although
we need the physical information only within the time range,
$z=0\sim 64$, it is impossible to get it without the duplication from
the time range $z=64\sim 128$.

Now that the most essential part of the dynamical description, i.e.,
Eq.~\eqref{eq:matrix_ele} is nicely reproduced, the vacuum expectation
value can be immediately derived from the convolution with the wave
function.  We do not show this part of the numerical results because
the calculation is just a Gaussian integral.  It is quite conceivable
that the same machinery should be effective even when the initial
state is not a simple Gaussian function like Eq.~\eqref{eq:gaussian}
(which is adopted here to yield the vacuum expectation value and was
also assumed in non-equilibrium study in Ref.~\cite{Berges:2006xc}
just for brevity).  For the extensive investigation of non-equilibrium
phenomena using the CTP formalism in stochastic quantization, we will
report our results in follow-up publication.  We shall close our
present discussions with this simple but clear-cut demonstration of
the strength of the CTP formalism.

\section{Summary and future extensions}
\label{sec:summary}

We have investigated the feasibility of stochastic quantization in a
simple system of 0+1 dimensional scalar theory or a quantum mechanical
system.  We focus on the vacuum properties with the $\rmi\epsilon$
prescription and have tested the convergence.  As long as the vacuum
properties are concerned, the boundary condition in time is irrelevant
thanks to the $\rmi\epsilon$ prescription, and we can work in
momentum-frequency space (that corresponds to the periodic boundary
condition).  We find it easier to enhance the numerical stability in
frequency space and have succeeded in performing the stable simulation
taking account of interaction effects.

Because we can alternatively solve 0+1 dimensional $\phi^4$ theory or
an anharmonic oscillator problem in quantum mechanics by
diagonalization of the Hamiltonian with sufficiently large number of
bases, we have made a quantitative comparison between the numerically
exact results and our results from stochastic quantization.  Although
the pole of the propagator or the effective mass behaves consistently
with the correct answer, there are unphysical width and residue
appearing in the propagator, which indicates that the numerical
solution of stochastic quantization falls into some unphysical
fixed-point.  We could give an argument on the existence of such
unphysical fixed-point using the alternative representation using the
Fokker-Planck equation.  We propose a prescription to overcome this
problem; that is, the restricted phase-space approximation (RPSA) that
would be exact in the $\mathrm{O}(N\to\infty)$ theory.  In the RPSA
the interaction is modified in such a way that the allowed phase space
is limited.  In frequency space, in particular, the RPSA makes it
possible to implement the interaction in a semi-local manner and to
improve the numerical stability drastically.  Our comparison has
revealed that the RPSA results are remarkably close to the mean-field
estimate of the effective mass, which are also close to the full exact
answer.  It should be an interesting future work to test the further
potential of the RPSA in 3+1 dimensional systems.  We have performed
some preliminary simulations and it is likely that the stable
simulation is feasible enough to have physically meaningful results.

It is not yet clear if the RPSA can describe general non-equilibrium
phenomena;  the RPSA becomes most effective when formulated in
momentum-frequency space but the periodicity can be lost for problems
out of the vacuum state.  Nevertheless, it is expected to work near
the thermal equilibrium; for example, we can apply our real-time
method to compute the spectral functions at finite temperature.  Also,
the particle production problem under time-dependent external fields
would be an ideal setup to test the merit of the RPSA;  this is a
phenomenon associated with the change of the ``vacuum'' induced by
external fields, which we can investigate not losing the advantage of
the $\rmi\epsilon$ prescription~\cite{Fukushima:2014iqa}.

On the formal level, as we already mentioned, the RPSA would become
exact in the large-$N$ limit.  It should be theoretically an
intriguing question to formulate the $\mathrm{O}(N\to\infty)$ theory
with stochastic quantization, which may provide us with a hint to
represent the theory in higher dimensions (with an extra coordinate of
the fictitious time added).  It would be conceivable that stochastic
approaches could be useful to deepen our understanding on the
holographic duality between classical and quantum theories.

We are now proceeding to the application of stochastic quantization
for fully non-equilibrium phenomena.  As a preparation for this, we
have presented an explicit check of the closed-time path (CTP)
formalism for the non-interacting case.  The time evolution of the
expectation value of an operator is correctly reproduced from the
initial time $\ti$ to the final time $\tf$ and it is reflected at
$t=\tf$ that separates the forward path and the backward path.  This
indicates that the Feynman kernel is correctly calculable, and in
principle, the time-dependence starting with arbitrary initial
condition would be available from the convolution with the initial
wave function.  We are making progress in this direction including
interaction effects.

\section*{Acknowledgments}
K.~F.\ would like to thank J\"{u}rgen~Berges, Jan~Pawlowski, and
D\'{e}nes~Sexty for useful conversations.  K.~F., Y.~H., and
T.~O.\ were partially supported by JSPS KAKENHI Grant Number 24740169,
24740184, and 23740260, respectively.  Y.~H. was also supported by the
RIKEN iTHES Project.

\appendix

\section{One-loop self-energy calculation}
\label{app:oneloop}

We explicitly check if Eq.~\eqref{eq:average} holds for the one-loop
self-energy.  For notational convenience we define the inverse free
propagator as
\begin{equation}
 \rmi\,\Delta^{-1} \equiv \omega^2 - \xi_{\bk}^2 + \rmi \epsilon\;.
\end{equation}
Then, using this notation, we can express the two-point function order
by order in terms of the coupling $\lambda$.  In the leading order
(i.e., zeroth order in $\lambda$ referred by a superscript $(0)$
here), the left-hand side of Eq.~\eqref{eq:average} gives the free
propagator by definition.  The right-hand side takes a more
non-trivial form that is
\begin{equation}
 \overline{\langle\phi_k(\theta)\,\phi_{k'}(\theta)\rangle_\eta^{(0)}}
  = (2\pi)^d\,\delta(k+k')\,\Delta(k)\,\frac{1}{\theta}
    \int_0^\theta \rmd\theta'\, \bigl[ 1-\rme^{-2\theta'\Delta^{-1}(k)}
    \bigr] \;.
\end{equation}
We can easily perform the $\theta$-integration in the above expression
to find that the free propagator (that is what our calculation is
supposed to get) is multiplied by an extra factor,
$1+(\Delta/\theta)(\rme^{-2\theta\Delta^{-1}}-1)/2$.  The modulus of
the deviation from the unity is now given by
\begin{equation}
 \Bigl| \frac{\Delta(k)}{\theta}\cdot
  \frac{\rme^{-2\theta\Delta^{-1}(k)}-1}{2} \Bigr| \;\le\;
  \Bigl| \frac{\Delta(k)}{\theta} \Bigr| \cdot
  \Bigl| \frac{\rme^{-2\theta\Delta^{-1}(k)}-1}{2} \Bigr| \;\le\;
  \Bigl| \frac{\Delta(k)}{\theta} \Bigr| \;\le\;
  \frac{1}{|\theta\epsilon|} \;.
\end{equation}
Sending $\theta$ to infinity while keeping a small but finite
$\epsilon$, we can safely drop this extra term and we can recover the
free propagator as we should.

Now let us go to the next order that contributes to the one-loop
diagram of the self-energy.  Up to the first order in $\lambda$
(referred by $(1)$ here), we can perform tedious but straightforward
calculations to reach eventually the following expression,
\begin{align}
 & \langle\phi_k(\theta)\,\phi_{k'}(\theta)\rangle_\eta^{(1)}
  = -3\rmi\lambda\,(2\pi)^d\,\delta^{(d)}(k+k')\,\Delta^2(k)
  \int\frac{\rmd^d k_1}{(2\pi)^d}\,\Delta(k_1)\,\biggl[ 1
  -\rme^{-2\theta\Delta^{-1}(k)} \notag\\
 &\qquad\qquad -2\rme^{-2\theta\Delta^{-1}(k)}\, \theta \Delta^{-1}(k)
   -\frac{1}{1-\Delta(k)\Delta^{-1}(k_1)}\bigl(
  \rme^{-2\theta\Delta^{-1}(k_1)}-\rme^{-2\theta\Delta^{-1}(k)} \bigr)
  \notag\\
 &\qquad\qquad\qquad\qquad
  -\Delta^{-1}(k)\,\Delta(k_1)\,\rme^{-2\theta\Delta^{-1}(k)}
  \bigl( \rme^{-2\theta\Delta^{-1}(k_1)} -1 \bigr) \biggr] \;.
\end{align}
This complicated expression reduces to the standard expression of the
self-energy once we take the $\theta\to\infty$ limit.  Then, we can
drop $\rme^{-2\theta\Delta^{-1}(k)}$ from the above and we correctly
reproduce,
\begin{equation}
 \lim_{\theta\to\infty} \langle\phi_k(\theta)\,\phi_{k'}(\theta)
  \rangle_\eta^{(1)} = (2\pi)^d\,\delta^{(d)}(k+k')\,
  \Delta^2(k)\,(-3\rmi\lambda)\int\frac{\rmd^d k_1}{(2\pi)^d}
  \,\Delta(k_1) \;.
\label{eq:selfenergy}
\end{equation}
In the same way as the previous example for the free propagator we can
proceed to the $\theta$-averaged calculation.  The final results read,
\begin{align}
 & \overline{\langle\phi_k(\theta)\,\phi_{k'}(\theta)\rangle_\eta^{(1)}}
  = (2\pi)^d\,\delta^{(d)}(k+k')\,\Delta^2(k)\,(-3\rmi\lambda)
  \int\frac{\rmd^d k_1}{(2\pi)^d}\,\Delta(k_1) \notag\\
 &\qquad \times \biggl[ 1+\rme^{-2\theta\Delta^{-1}(k)}
  -\frac{\Delta(k)}{\theta}\bigl( 1-\rme^{-2\theta\Delta^{-1}(k)}
  \bigr) - \frac{1}{1-\Delta(k)\Delta^{-1}(k_1)} \notag\\
 &\qquad\quad \times \Bigl(\frac{\Delta(k_1)}{\theta}\cdot
  \frac{1-\rme^{-2\theta\Delta^{-1}(k_1)}}{2} - \frac{\Delta(k)}{\theta}
  \cdot\frac{1-\rme^{-2\theta\Delta^{-1}(k)}}{2} \Bigr) \notag\\
 &\qquad\quad
   -\Delta^{-1}(k)\Delta(k_1) \Bigl( \frac{1}{\theta(\Delta^{-1}(k)
  +\Delta^{-1}(k_1))}\cdot\frac{1-\rme^{-2\theta(\Delta^{-1}(k)
  +\Delta^{-1}(k_1))}}{2} \notag\\
 &\qquad\qquad -\frac{\Delta(k)}{\theta}\cdot
  \frac{1-\rme^{-2\theta\Delta^{-1}(k)}}{2} \Bigr) \biggr] \;.
\end{align}
Using the same inequality we can soon confirm that all additional
terms in the square brackets are vanishing in the limit of
$\theta\to\infty$ and then the above complicated expression simplifies
to the standard one in Eq.~\eqref{eq:selfenergy}.

\section{Diagonalization of the Hamiltonian in 0+1 dimensions}
\label{app:full}

The one-dimensional Hamiltonian reduces to an as simple form as
\begin{equation}
 H = \frac{\pi^2}{2}+\frac{\phi^2}{2} +\frac{\lambda}{4} \phi^4 \;,
\label{eq:simpleH}
\end{equation}
where $m=1$ and the commutation relation is $[\phi,\pi]=\rmi$.  This
is a problem of quantum mechanics, which is numerically solvable by
diagonalizing the Hamiltonian.  Here, we introduce the
annihilation/creation operators as
\begin{equation}
 a = \frac{1}{\sqrt{2}}(\phi+i\pi) \;, \qquad
 a^\dag = \frac{1}{\sqrt{2}}(\phi-i\pi) \;,
\end{equation}
which satisfy $[a,a^\dag]=1$.  The harmonic part is
${\pi^2}/2+{\phi^2}/2=N+1/2$ with the number operator $N=a^\dag a$.
Using $[N,a]=-a$ it is easy to show $[N,a^\dag]=a^\dag$,
$a^2a^{\dag 2}=(N+2)(N+1)$, and $a^{\dag2}a^2=N(N-1)$.  Then we can
expand the $\phi^4$ term as
\begin{align}
 \phi^4 &= \frac{1}{4}(a+a^\dag)^4 \notag\\
 &= \frac{1}{4}\bigl[ a^4+a^{\dag 4}+6N^2+6N+3 +2a^2(2N-1)
  +2a^{\dag 2}(2N+3) \bigr]\;.
\end{align}
We utilize the eigenvalue bases of $N$; i.e., $|n\rangle$, which we
can express as $|n\rangle \equiv (a^\dag)^n|0\rangle/\sqrt{n!}$ using
the creation operators.  The matrix element of $\phi^4$ is
\begin{equation}
 \begin{split}
 \langle n|\phi^4|m \rangle &=
  \frac{1}{4} \delta_{n,m} (6m^2+6m+3)
   + \frac{1}{2}\delta_{n+2,m}\sqrt{(n+2)(n+1)}(2n+3) \\
 &\qquad + \frac{1}{2}\delta_{n,m+2}\sqrt{(m+2)(m+1)}(2m+3) \\
 &\qquad + \frac{1}{4}\delta_{n,m+4}\sqrt{(m+4)(m+3)(m+2)(m+1)} \\
 &\qquad + \frac{1}{4}\delta_{n+4,m}\sqrt{(n+4)(n+3)(n+2)(n+1)} \;.
 \end{split}
\end{equation}
Therefore, the matrix element of the Hamiltonian~\eqref{eq:simpleH} is
\begin{equation}
 \begin{split}
 \langle n|H|m\rangle &=
  \delta_{n,m} \biggl[ m+\frac{1}{2}+\frac{\lambda}{16}
   (6m^2+6m+3) \biggr] \\
 &\qquad + \frac{\lambda}{8}\delta_{n+2,m}\sqrt{(n+2)(n+1)}(2n+3) \\
 &\qquad + \frac{\lambda}{8}\delta_{n,m+2}\sqrt{(m+2)(m+1)}(2m+3) \\
 &\qquad + \frac{\lambda}{16}\delta_{n,m+4}\sqrt{(m+4)(m+3)(m+2)(m+1)}\\
 &\qquad + \frac{\lambda}{16}\delta_{n+4,m}\sqrt{(n+4)(n+3)(n+2)(n+1)} \;.
 \end{split}
\end{equation}
We can obtain the propagator in momentum space as
\begin{align}
 G(\omega) &= \int \rmd t\, \rme^{\rmi\omega t}
  \langle T\phi(t)\phi(0) \rangle \notag\\
 &= \int \rmd t\, \rme^{\rmi\omega t} \theta(t)
  \langle \phi(t)\phi(0) \rangle
  +\theta(-t)\langle \phi(0)\phi(t) \rangle \notag\\
 &= \sum_{n}\frac{2\rmi\delta E_n}{\omega^2-(\delta E_n)^2} \bigl|
  \langle E_n|\phi(0)|E_0 \rangle \bigr|^2 \;,
\label{eq:propagator}
\end{align}
where $|E_n \rangle$ is the eigenstate of $H$ with the energy
eigenvalue $E_n$.  We also introduced a notation,
$\delta E_n\equiv E_n-E_0$.  The matrix element
$\langle E_n|\phi(0)|E_0 \rangle$ is expressed as
\begin{align}
 & \langle E_n|\phi(0)|E_0 \rangle = \sum_{m,l}
  \langle E_n| m\rangle\langle m|\phi|l\rangle \langle l|E_0 \rangle
  \notag\\
 &\qquad\qquad= \sum_{m,l} \frac{1}{\sqrt{2}}\langle E_n|m\rangle
  \langle m|a+a^\dag|l\rangle \langle l |E_0 \rangle \notag\\
 &\qquad\qquad= \sum_{m,l} \frac{1}{\sqrt{2}}\langle E_n|m\rangle
 \langle l|E_0 \rangle \bigl( \delta_{m+1,l}\sqrt{m+1} +
 \delta_{m,l+1}\sqrt{l+1} \bigr) \notag\\
 &\qquad\qquad= \sum_{m} \sqrt{\frac{m+1}{2}}
  \bigl( \langle E_n| m\rangle \langle m+1 |E_0 \rangle +
  \langle E_n| m+1\rangle \langle m |E_0 \rangle \bigr) \;.
\end{align}
We use the above form for the numerical calculation, which quickly
converges to the exact answer.  We can confirm that the above reduces
to the free expression in the case of $\lambda=0$ and thus
$|E_n\rangle=|n\rangle$.  Plugging
$\langle E_n|\phi(0)|E_0 \rangle=\delta_{n,1}/\sqrt{2}$ and
$\delta E_n=n$ into the above, we can arrive at the following
expression as
\begin{equation}
 G(\omega) = \frac{\rmi}{\omega^2 -1} \;.
\end{equation}
Here we note that we use a unit with the mass $m=1$ and if we retrieve
the mass explicitly, the denominator is given by $\omega^2-m^2$ in the
above.

\bibliographystyle{utphys}
\bibliography{stochastic}

\end{document}